# Front Matter

(scroll down)

**Contribution to Topical Issue of Solar Physics on
"Recalibration of the Sunspot Number"**

**"We solicit manuscripts on this general subject for inclusion in a Topical Issue (TI) of Solar Physics. The deadline for submission of statements of interest (SOI) with a tentative title, author list, and three suggestions for referees is 15 June 2015, and the deadline for manuscript submission is 31 October 2015."**

**This Topical Issue is an outgrowth of the Sunspot Number Workshop Series (ssnworkshop.wikia.com/wiki/Home). Workshops were held at the National Solar Observatory in Sunspot, New Mexico (September, 2011), the Royal Observatory of Belgium in Brussels (May, 2012), the National Solar Observatory in Tucson, Arizona (January, 2013), and Specola Solare Ticinese in Locarno, Switzerland (May 2014). This Topical Issue is not a conference proceedings, and it is not limited to research presented at the workshop. Nor is it limited to scientists who participated in any of the workshops. All submissions must be completed original papers that meet the regular quality of the Journal. The Topical Issue will begin with several invited reviews to summarize the subject and frame the work in the research papers which follow.**

# Reconstruction of the Sunspot Group Number: the Backbone Method

**Leif Svalgaard[1] and Kenneth H. Schatten[2]**

[1] Leif Svalgaard, leif@leif.org
HEPL, Stanford University, Stanford, CA 94305, USA

[2] Kennuth H. Schatten
a.i. solutions, Suite 215, Derekwood Lane, Lanham, MD 20706, USA

**Abstract** We have reconstructed the sunspot group count, not by comparisons with other reconstructions and correcting those where they were deemed to be deficient, but by a re-assessment of original sources. The resulting series is a pure solar index and does not rely on input from other proxies, *e.g.* radionuclides, auroral sightings, or geomagnetic records. 'Backboning' the data sets, our chosen method, provides substance and rigidity by using long-time observers as a stiffness character. Solar activity, as defined by the Group Number, appears to reach and sustain for extended intervals of time the same level in each of the last three centuries since 1700 and the past several decades do not seem to have been exceptionally active, contrary to what is often claimed.

## 1. Introduction

Reports of the appearance of sunspots go back to antiquity, but the *sunspot record* in the Western World begins in and about 1610 with the first telescopic observations of sunspots by Harriot, Scheiner, Galileo, and others and continues to the present day (*e.g.* Vaquero and Vázquez, 2009). Rudolf Wolf collected much of the earlier observations and provided 45 years of observations of his own, constructing his celebrated Relative Sunspot Number (SSN), which we still use today (http://www.sidc.be/silso/home). As the small spots were more difficult to observe with telescope technology of the 17th and early 18th centuries (*e.g.* with chromatic and spherical aberration) Hoyt and Schatten (1998, hereafter HS) suggested using instead the count of sunspot groups (active regions) as a measure of solar activity on the notion that groups were easier to see and fewer would be missed. Hoyt and Schatten performed a very valuable service by finding and digitizing many sunspot observations not known or used by Wolf and his successors, effectively doubling the amount of data available before Wolf's tabulations. The resulting Group Sunspot Number (GSN) was made compatible with the SSN by scaling the number of groups (GN) by a factor (12.08) to make it match the modern SSN. Unfortunately, the two series disagree seriously before ~1885 and the GSN has not been maintained after the 1998 publication of the series.

### 1.1. The Sunspot Number Workshops

The Sunspot Number Workshops (http://ssnworkshop.wikia.com/wiki/Home; Cliver *et al.*, 2015) were convened with the goal of reconciling the two series and to provide the community with an up-to-date, unified and thoroughly vetted series for correlative studies. In the words of Jan Stenflo, http://www.leif.org/research/SSN/Stenflo.pdf, we can make an "analogy with the cosmic distance scale: One needs a ladder of widely different techniques valid in a sequence of partially overlapping regimes. Similarly, to explore the history of solar variability we need a ladder of overlapping regimes that connect the present physical parameters (TSI, magnetograms, F10.7 flux, UV radiance, etc.) with the distant past. The time scale from the present back to Galileo can only be bridged by the Sunspot Number, which in turn allows the ladder to be continued by isotope methods, etc". The review paper by Clette *et al.* (2014) summarizes the work and conclusions of the Workshops. The review paper referred to the *backbone-method* for constructing the GN time series. The present paper constitutes publication in the open literature of the details of that method, and should be viewed as a step on the road to the long-needed and epoch-making recalibration of the venerable sunspot number 'distance scale' (Cliver *et al.*, 2013, Clette *et al.*, 2014).

## 2. The Data

The basic data are the listing of sunspot groups by year and by observer published by HS and held at the National Centers for Environmental Information (NCEI, now also holding data from what was formerly the National Geophysical Data Center NGDC) http://www.ngdc.noaa.gov/stp/space-weather/solar-data/solar-indices/sunspot-numbers/group/. Unfortunately, the links are not stable so the data are also duplicated in the database forming part of the supplementary data of the present paper. Research since HS has revealed (and corrected) problems, inconsistencies, and mistakes in the HS database (*e.g.* Vaquero and Trigo, 2014, Vaquero *et al.*, 2011, Vaquero *et al.*, 2015, Willis *et al.*, 2013) as well as added more data, both in the early part and in the time frame after ~1945. Most of the time, HS did not recount the groups from original sources, but instead utilized counts published by others, typically by Wolf. Most of our knowledge about sunspots in the 18[th] century relies on sunspot drawings by J.C. Staudach (digitized by Arlt, 2008). A detailed account of all these corrections and updates is beyond the scope of the present paper, but is a very important part of the patrimonium that we seek to safeguard and will be presented elsewhere.

### 2.1. Averaging

The individual observations, if any, for each observer are first used to compute monthly averages for that observer. A yearly average for the observer is then calculated from the average of all months of the year in which there was at least one observation. This minimizes the effect of clustering in time, but also introduces noise stemming from months with only a few observations. As HS point out, only 5% coverage is enough to yield a reliable yearly average. At any rate, for times that are data-rich, no distortions are introduced by our averaging scheme, and for times that are data-poor, there is not much else we can do to improve matters. HS had employed a special 'fill-in' procedure for days with no observations in order to 'bridge' data gaps, presumably to improve the statistics. We shall not do that directly, although using a monthly mean may be considered as filling in days with no data with the mean of days with data.

## 2.2. Problems Comparing Observers

In comparing the counts from one observer to those by another the ideal would be to compare the counts on days where both observers have made a definite observation. This will work at data-rich times, but fails at data-poor times, as there may be very few days in a year, if any at all, where we have simultaneous observations. Our averaging scheme overcomes the problem of lack of direct overlap, relying on the high degree of autocorrelation of solar activity, but, of course, introduces extra variance. Simulations, 'thinning' data-rich observing sequences, show that the error introduced by this is but minor (less than 10% for up to 85% thinning). In the final analysis, one cannot 'make up' data out of whole cloth. So statistics will remain poor unless there are real data to work with. A similar problem arises if you, as HS, only compare days where both observers report a non-zero count. This introduces an undesirable bias if one observer (e.g. with a smaller telescope) reports many more days without spots than the other, as well as decreasing the number of days to be compared. Realizing that no method is perfect, we welcome workable suggestions to overcome those difficulties. At this point we go with the simple averaging scheme described above.

## 2.3. Know Thy Data

Figure 1 shows the average number of groups per year (blue curve) since 1610 and shows (on a logarithmic scale, green curve) the number of observations per year by all observers.

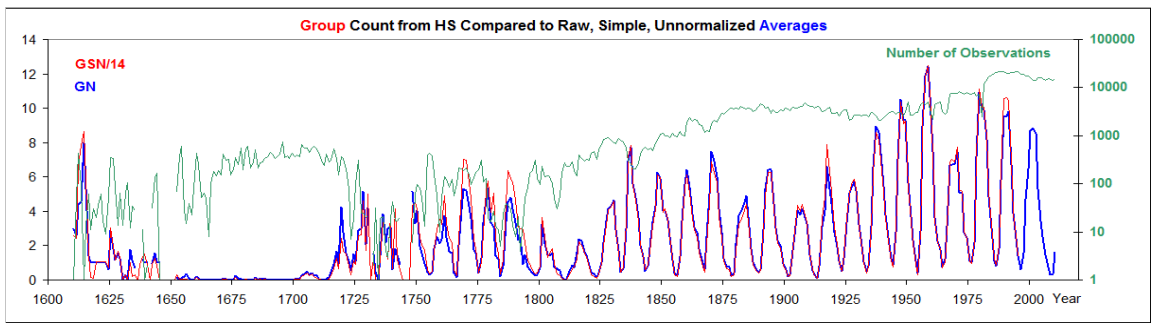

Figure 1: Average number of groups per year with no normalization (blue curve) overlain (red curve) by the Group Sunspot Number from HS divided by 14 (to have the same mean as just the number of groups). Additionally, the total number of observations by all observers for each year is shown on a logarithmic scale (green curve; no points are plotted for years with no observations).

It is remarkable that the raw data with no normalization at all closely match (coefficient of determination for linear regression $R^2 = 0.97$) the number of groups calculated by dividing the HS GSN by an appropriate scale factor (14), demonstrating that the elaborate, and somewhat obscure and likely incorrect, normalization procedures employed by HS have almost no effect on the result. The normalization thus did not introduce or remove any trends (such as the 'secular increase' from 1700 to the present) or anomalies that were not already in the raw data, contrary to the inference made by Cliver and Ling (2015).

# 3. The Backbones

Building a long time series from observations made over time by several observers can be done in two ways: *Daisy-chaining*: successively joining observers to the 'end' of the series, based on overlap with the series as it extends so far (accumulates errors), and *Backboning*: find a primary observer for a certain (long) interval and normalize all other observers individually to the primary based on overlap with only the primary (no accumulation of errors). The selection of the primary observer should be based both on the length of the observational series (as long as possible) as on the perceived 'quality' of the observations such as regularity of observing, suitable telescope, and lack of obvious problems. We start by constructing two backbones, the Schwabe (1794-1883) and the Wolfer (1841-1944) backbone. The Schwabe backbone is centered on the observing interval for Schwabe and includes all 'reliable' observers who overlap in time with Schwabe. The reliability is judged by how high the correlation is between yearly averages of the observations by the observer and by Schwabe. Similarly, the Wolfer backbone includes all reliable observers who overlap with Wolfer. The two backbones overlap by 42 years so can be cross-calibrated with confidence. Figure 2 gives an overview of the time intervals observed by the observers listed.

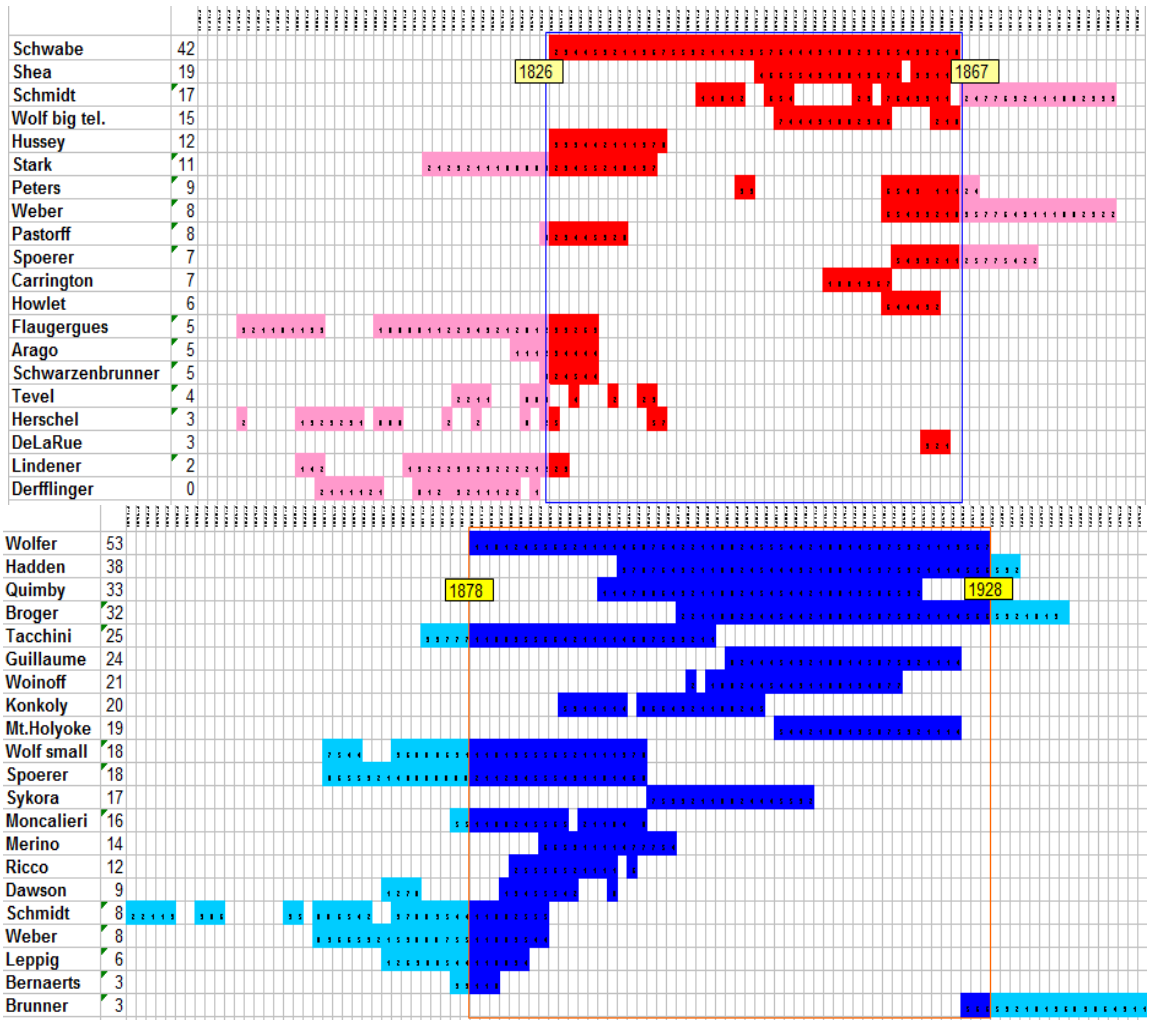

Figure 2: (Top) Coverage and observers for the Schwabe Backbone (1794-1883). (Bottom) Coverage and observers for the Wolfer Backbone (1841-1944). The numbers next to the observer names signify the number of years of overlap with the primary observer.

For each Backbone, we regress the primary observer's group count against each observer's count for each year and we plot the result (for examples see Figure 3; the supplementary data contains the plots for all observers used in this study). Experience shows that the regression line almost always very nearly goes through the origin, so we force it to do so and calculate the slope and various statistics, such as 1-σ uncertainty and the *F*-value, the ratio between the variance 'explained' by the fit and the residual unexplained variance. The comparison plots show that the data usually have homoscedasticity (roughly constant variance for all values of the independent variable), guarding against overestimating the goodness of fit.

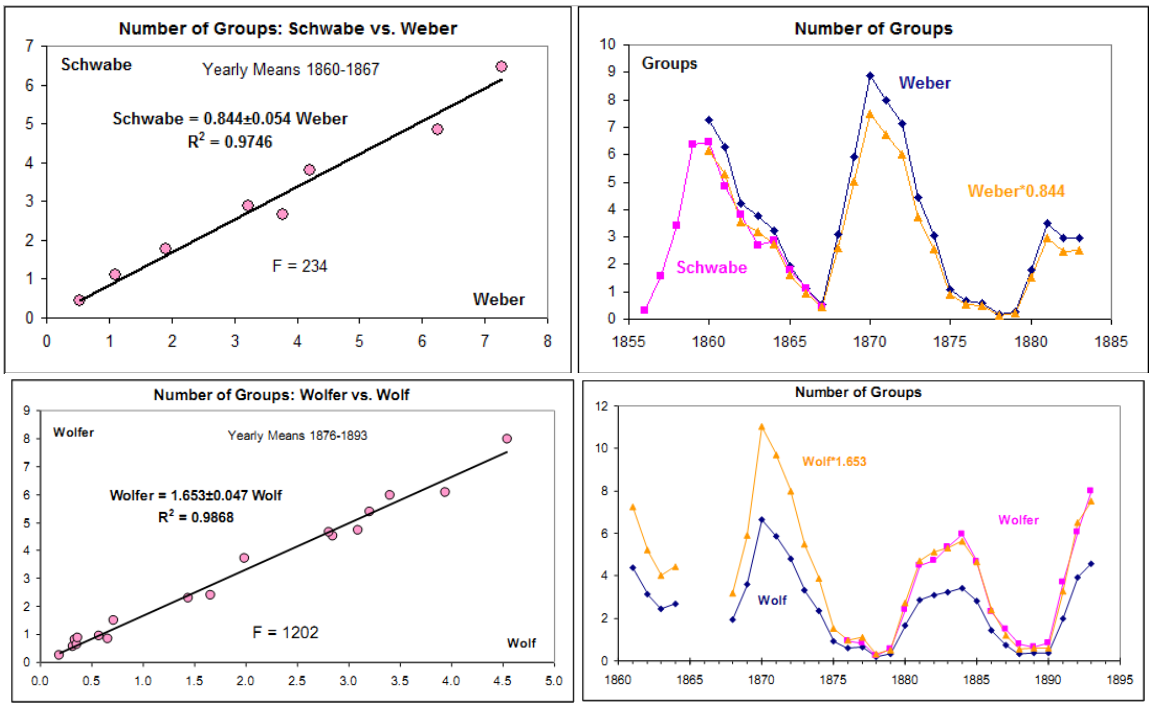

Figure 3: (Top) Regression of the number of groups observed 1860-1867 by Schwabe against the number of groups observed by Weber in Peckeloh. (Bottom) Regression of number of groups observed 1876-1893 by Wolfer (with standard 80 mm aperture telescope) against the number of groups observed by Wolf (with small 37 mm aperture telescope).

The slope of the regression gives us the factor by which to multiply the observer's count by to match the primary's count. The right-hand graph in the lower panel shows the result for the Wolfer Backbone: blue is Wolf's count (with his small telescope), pink is Wolfer's count (with the larger telescope), and the orange curve is the blue curve multiplied by the slope, bringing Wolf's observations on the same scale as Wolfer's. It is clear that the harmonization works well and that it shows that Wolfer with the larger telescope saw 65% more groups than Wolf did with the small, handheld telescopes (Figure 4) as one might expect.

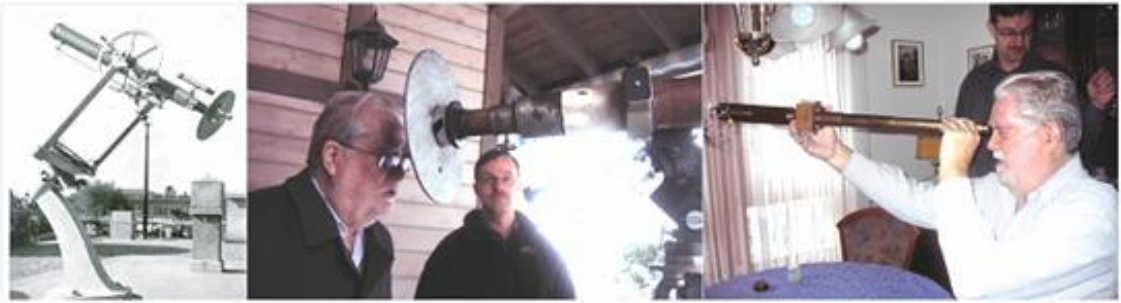

Figure 4: (Left) The 80 mm aperture (magnification X64) refractor used mostly by Wolf's assistants at the Zürich Observatory since 1863, designed by Joseph Fraunhofer and manufactured in 1822 at the Fraunhofer factory by his assistent Georg Merz. (Middle) The telescope still exists and is being used daily by Thomas Friedli (person at center). (Right) One of several small, portable, handheld telescopes (~40 mm aperture, magnification X20) used by Wolf almost exclusively from 1859 on, and is still in occasional use today.

It is noteworthy that the $k$-factor (used to normalize one observer to another, in this case to the Royal Greenwich Observatory, RGO) determined by HS for Wolf and Wolfer differ by only 2% while the straightforward comparison (Figure 3) clearly shows a 65% difference. We have not been able to reproduce the HS $k$-factors for secondary observers (those not directly comparing with RGO) using the information given in HS. In fact, there seems to be no correlation between the slopes of our straightforward regressions and the $k$-factors reported by HS. The reason for this is not known and is lost in time.

### 3.1. The Schwabe and Wolfer Backbones

Applying this methodology yields the two backbones (Figures 5 and 6). We stress that the backbones are independent and are based purely on solar observations with no empirical or *ad hoc* adjustments apart from the (necessary) harmonization just described.

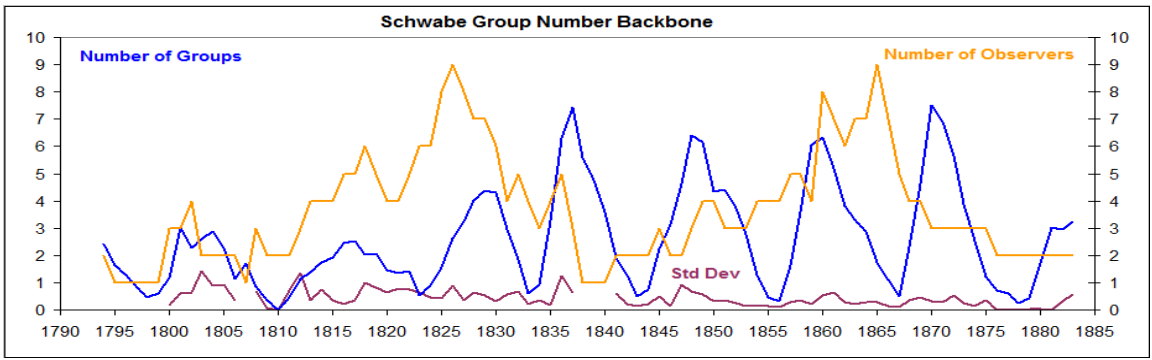

Figure 5: The Schwabe backbone (blue) computed as the average of the normalized counts for the observers listed in Figure 2. The number of observers

(orange) is also shown. The standard deviation of the values going into the average is shown at the bottom of the plot (purple).

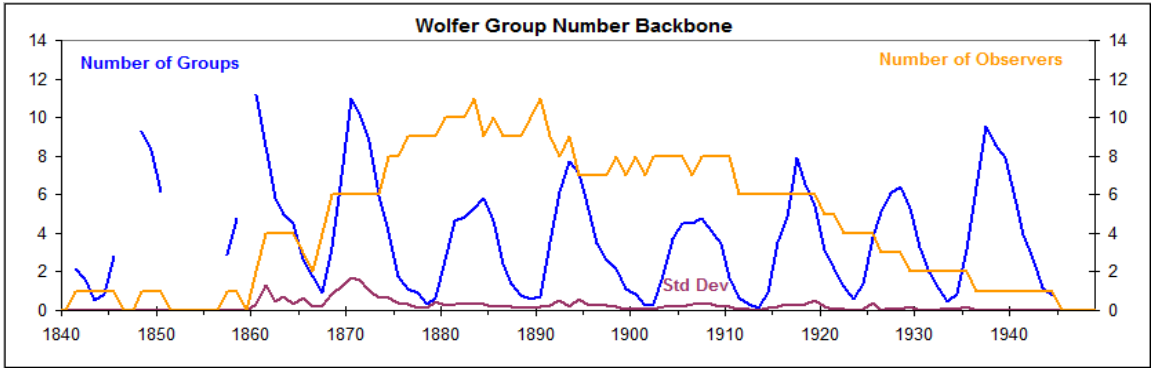

Figure 6: The Wolfer backbone (blue) computed as the average of the normalized counts for the observers listed in Figure 2. The number of observers (orange) is also shown. The standard deviation of the values going into the average is shown at the bottom of the plot (purple). The values before 1860 are based on Schmidt's observations and are uncertain and will be used with half weight.

The next order of business is to harmonize the two backbones, i.e. bring them onto the same scale. We shall use the Wolfer scale as the base scale, because of its larger number of (better documented) observers and choose the common interval 1861-1883 as the basis for the normalization (Figure 7).

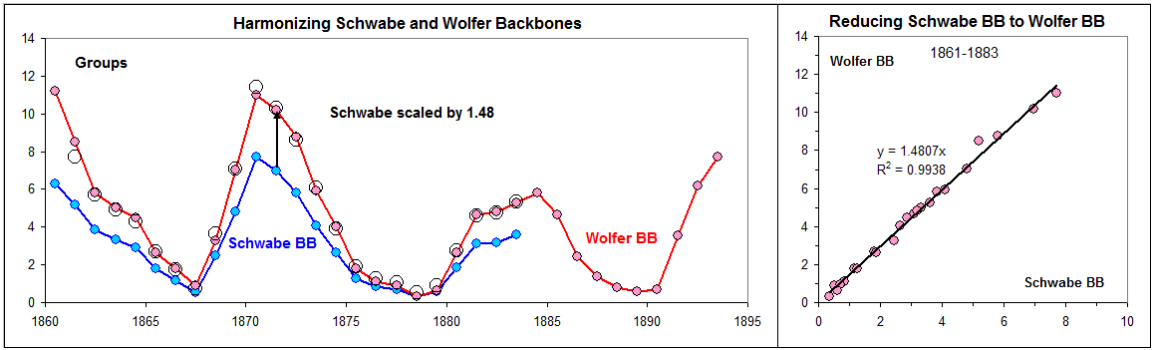

Figure 7: (Left) overlapping backbones: Wolf (blue) and Schwabe (red). Open circles show the result of scaling the Schwabe backbone upwards by a factor 1.48. (Right) Linear correlation showing the least-squares fit for the interval 1861-1883.

The scale factor 1.48±0.03 brings the Schwabe backbone onto the same scale as the Wolfer backbone, and 'explains' 99.4% of the variations of the two backbones, with no clear systematic variation with time, we can thus produce a composite series by multiplying the Schwabe backbone values by 1.48 and then simply average the resulting, normalized Schwabe backbone and the Wolfer backbone, giving half weight to the Schmidt data before 1861 (Figure 8).

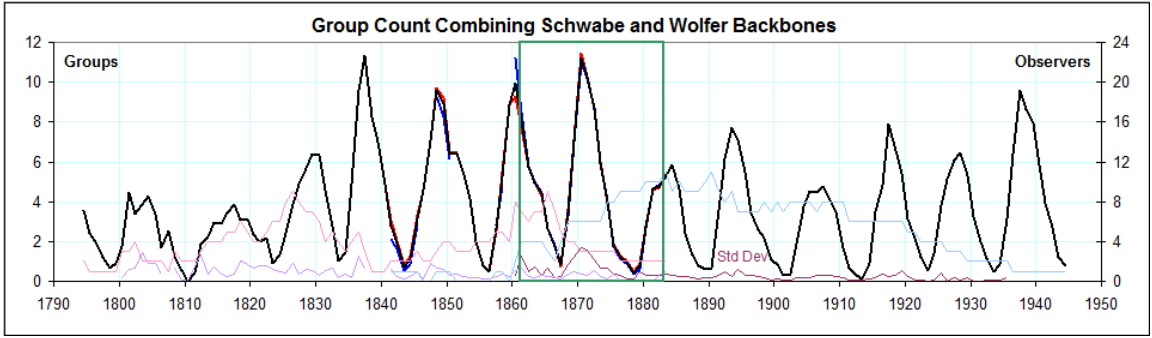

Figure 8: Composite, weighted-mean backbone 1794-1944, with the standard deviations of the two series (purple curves at the bottom of the Figure). The number of observers for the Schwabe backbone (itself shown by an underlying red curve) is shown as a pink curve, while the light blue shows the number of observers for the Wolfer backbone (itself shown by the blue curve), both using the right-hand scale. The green box shows the common interval used for the cross-calibration in Figure 7.

The counts for the years 1839 and 1840 are based only on Schwabe's observations and this presents a weakening of the confidence in the homogeneity of the backbone across those years. There is not much we can do about that at this point.

## 3.2. Comparison with the Wolf Sunspot Number

Although the relationship between the Group Number and Wolf's Relative Sunspot Number is not quite linear, partly because of the discontinuity between the sunspot numbers 0 (no spots) and 11 (one spot), it is possible and instructive to compare at least the higher values by scaling the Group Number by a suitable factor to match. Figure 9 shows a comparison for solar cycles 8 through 11, indicating that the SSN was reported ~20% too low for 1849-1863, similar to the reduction also found in a recent analysis by Leussu *et al.* (2013). It is significant that the Wolf sunspot numbers and the group numbers for cycle 8 are on par with the numbers for cycle 11 and that the Wolf numbers therefore have not been unduly inflated as suggested by Leussu *et al*.

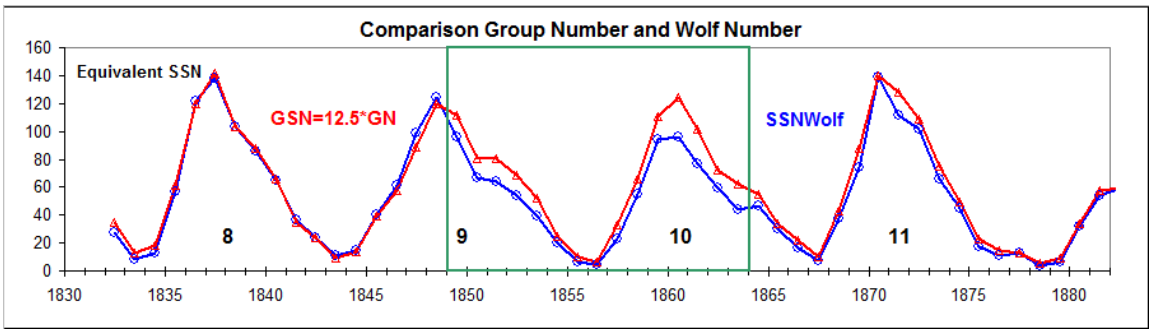

Figure 9: Comparison between the combined Schwabe-Wolfer backbone counts scaled by 12.5 (red curve) and the Wolf (International) Sunspot Number (blue curve). The scale factor was chosen to match the counts outside of the interval 1849-1863 (green box).

### 3.3. The Koyama and Locarno Backbones

We continue by constructing two further backbones reaching into the modern era, the Koyama (1920-1996) and the Locarno (1950-2015) backbones. We name the Koyama backbone in honor of the principal observer, Ms. Hisako Koyama, 小山 ヒサ子 (1916-1997). Ms. Koyama (Koyama, 1985) observed well into her eighties and detailed comparisons with other observers indicate a [not unexpected] decrease of visual acuity after ~1981, so we limit the regressions to the interval 1947-1980; note that this differs from the Koyama backbone reported in Clette *et al.* (2014). Figure 10 shows the coverage and observers in the same format as Figure 2:

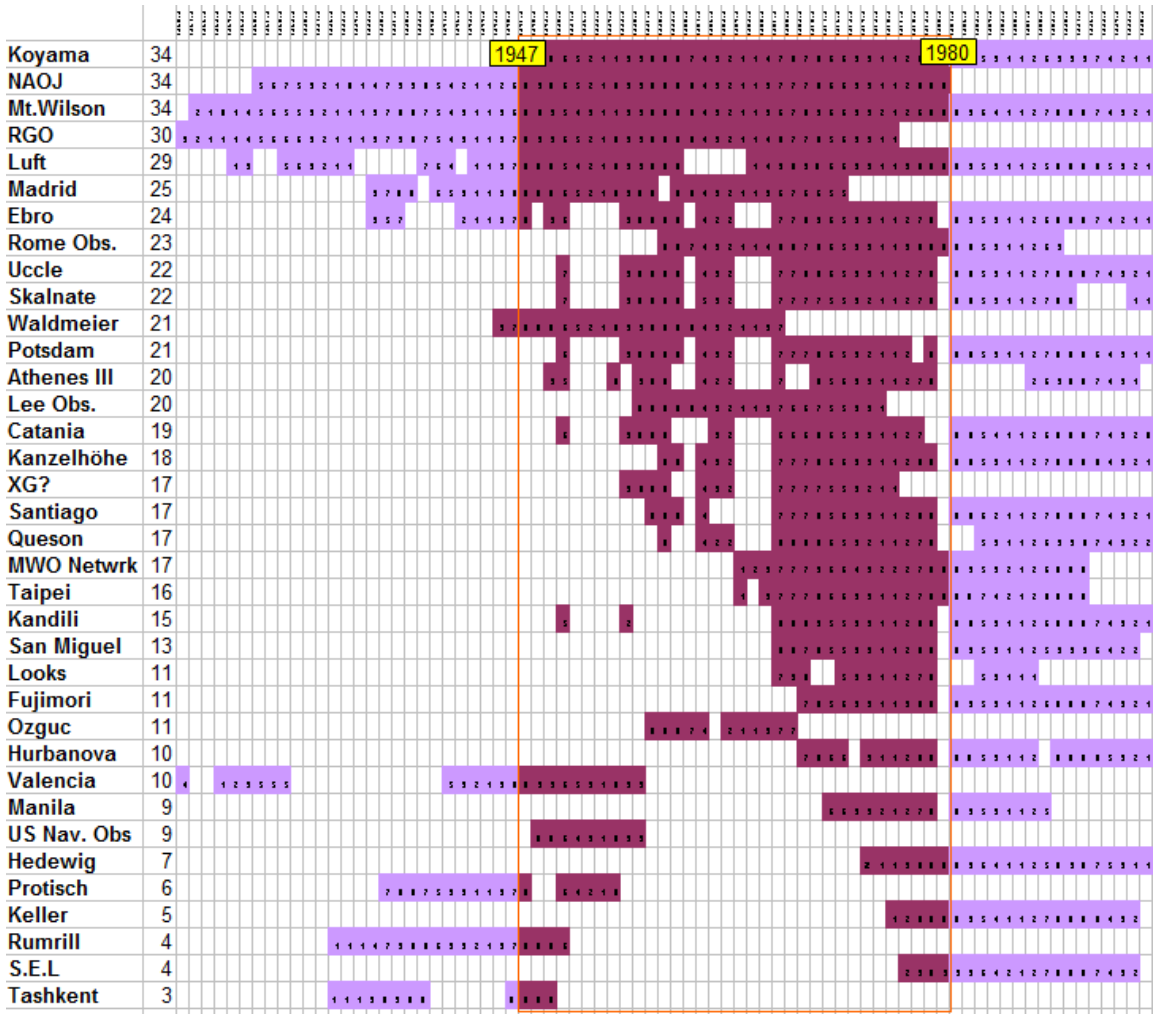

Figure 10: Coverage and observers for the Koyama backbone (1920-1996). The numbers next to the observer name signify the number of years of overlap with the primary observer, Koyama 1947-1980.

For the Koyama backbone we have included data from RGO which was not used in the Wolfer backbone because of the severe drift of the RGO group count before ~1915 (Cliver and Ling, 2015).

The resulting backbone is shown in Figure 11:

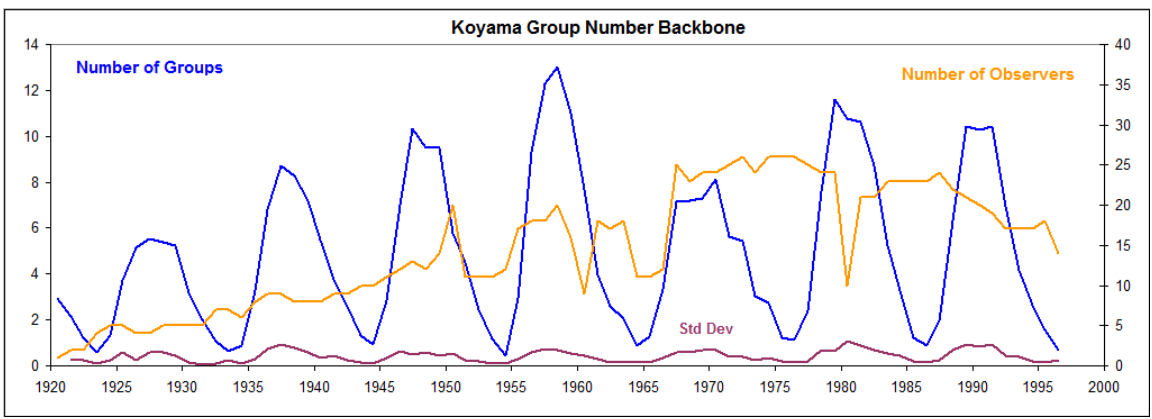

Figure 11: The new Koyama backbone (blue) computed as the average of the normalized counts for the observers listed in Figure 10. The number of observers (orange) is also shown. The sharp dip in the number of observers in the single year 1980 reflects the loss of records at the transition from Zürich to Brussels in that year. The standard deviation of the values going into the average is shown at the bottom of the plot (purple).

The standard deviation is almost a constant 8% of the group count, which then seems to be the inherent uncertainty or 'spread' from one ['modern'] observer to the next. So where the standard deviation is greater than that, like in Figures 5 and 6, the uncertainty is correspondingly greater. We can possibly use the solar cycle average standard deviation as a fraction of the solar cycle average of the group count for the cycle as a measure of the uncertainty for that cycle.

The Locarno station [on the shore of Lago Maggiore in Southern Switzerland] was established by Max Waldmeier in 1957 as an auxiliary observation site to Zürich to take advantage of the weather often being complementary opposite, at South and North of the Alps. After the termination of sunspot observations at Zürich, the continuation of the operation of the Locarno observatory (now named Specola Solare Ticinese) was guaranteed by a local association, the Associazione Specola Solare Ticinese, which was founded for that purpose. The sunspot data obtained at Specola serve up to now as the world reference for the sunspot number as the data from all other observers are normalized to the Locarno standard.

The primary observer at Locarno from the beginning up to the present has been Sergio Cortesi, observing using the same methods and procedures as at Zürich. This includes counting the sunspots visually at the eyepiece of a telescope stopped down to the same 80 mm aperture as that used in Zürich, and also following Waldmeier's prescription for weighting larger spots (up to five times) in the count, leading to an average inflation of 20% of the SSN. In addition, a drawing is made of the spots, organized into groups, visible on the solar disk. It is rare that the visual count is different from that reported on the drawing. We note that the inflation of the spot count due to the weighting does not carry over into the group count. Whether the modern group count likely is inflated due to a better understanding of what makes a group (after introduction of Waldmeier's classification for groups around 1938), as was suggested by the observer Zelenka (memorandum cited in Kopecký, 1980), will be addressed in section 4.

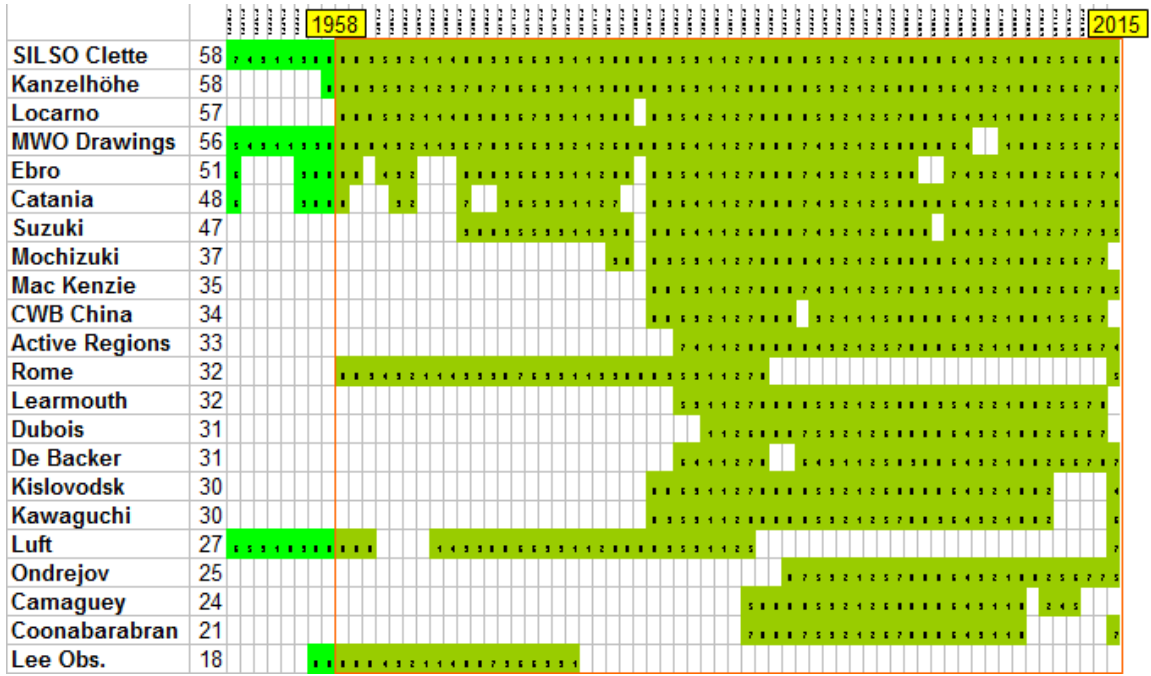

Figure 12: Coverage and observers for the Locarno backbone (1950-2015). The numbers next to the observer name signify the number of years of overlap with the primary observer, Locarno 1958-2015.

Figure 12 shows the coverage of data and observers used to construct the Locarno backbone, Figure 13:

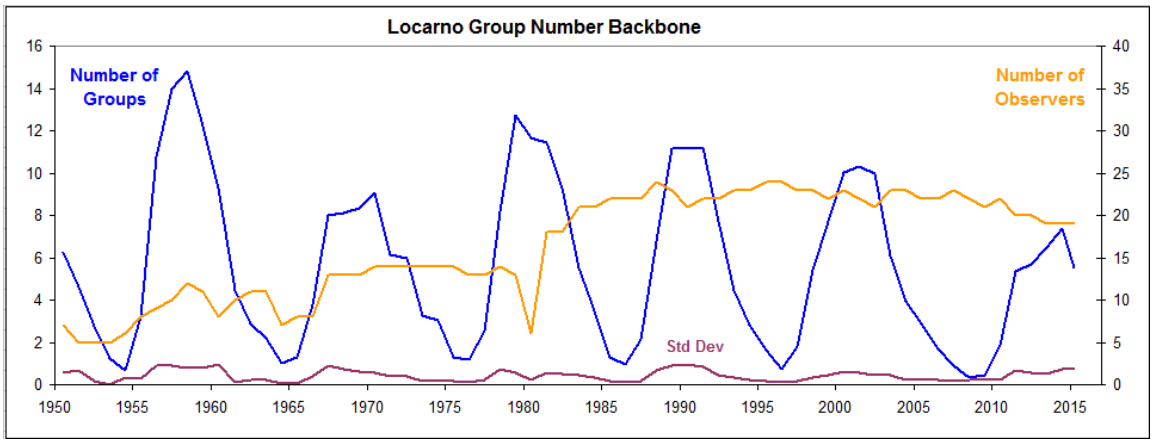

Figure 13: The Locarno backbone (blue) computed as the average of the normalized counts for the observers listed in Figure 12. The number of observers (orange) is also shown. The sharp dip in the number of observers in the single year 1980 reflects the loss of records at the transition from Zürich to Brussels in that year. The standard deviation of the values going into the average is shown at the bottom of the plot (purple).

The average standard deviation is, as for the Koyama backbone, the same 8%, which we, as before, interpret as the inherent uncertainty of the group count, stemming from the difficulty of apportioning crowded fields of multitudinous spots into groups. Contrary to common belief, counting spots is easy, counting groups is hard.

The next step is to harmonize (bring onto the same scale) the two backbones, see the middle panel of Figure 14:

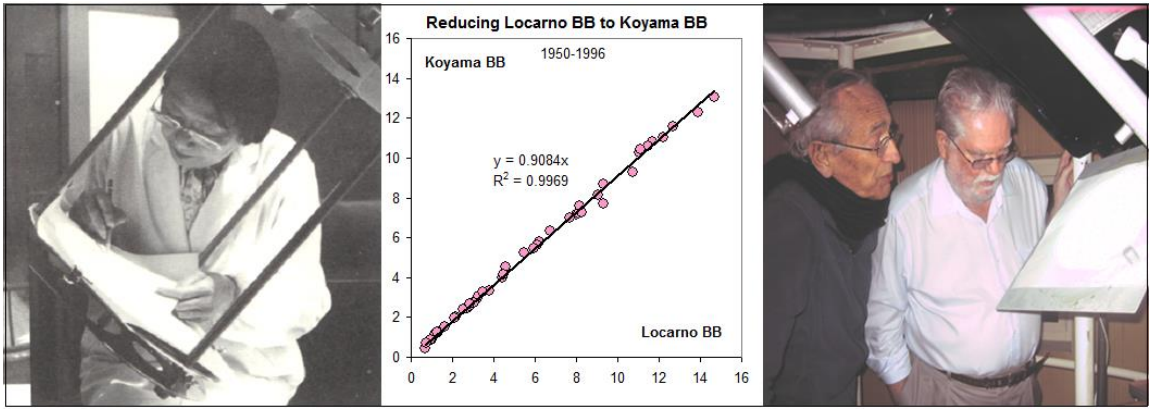

Figure 14: Hisako Koyama in Tokyo (left image) and Sergio Cortesi in Locarno (right image) at work. In a sense, Cortesi is the *real* keeper of the sunspot scale as SIDC (now renamed SILSO) normalized all other observers to Cortesi's count. The regression plot (middle) shows the relationship between the yearly average group counts for the Koyama and Locarno backbones.

The scale factor 0.91±0.01 brings the Locarno backbone onto the same scale as the Koyama backbone, and 'explains' 99.7% of the variations of the two backbones, with no significant systematic variation with time; we can thus produce a composite series by multiplying the Locarno backbone values by 0.91 and then average the resulting, normalized Locarno and Koyama backbones, Figure 15:

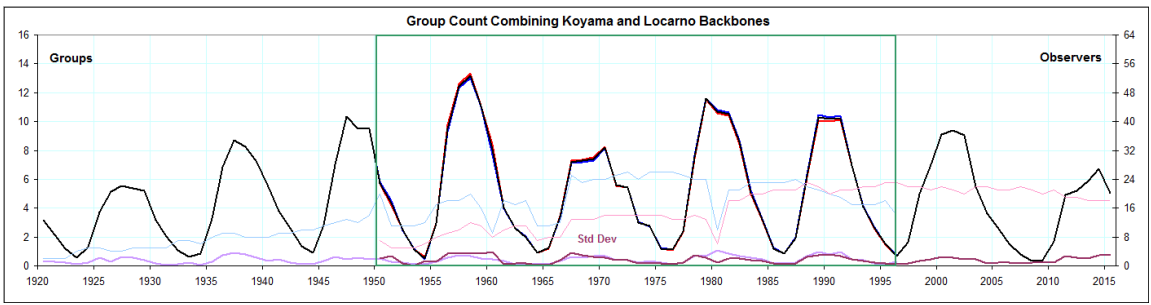

Figure 15: Composite, mean Koyama-Locarno backbone 1920-2015, with the standard deviations of the two series (purple curves at the bottom of the Figure). The number of observers for the Locarno backbone (underlying red curve) is shown as a thin pink curve (using the right-hand scale). The light blue curve (right-hand scale) shows the number for the Koyama backbone (underlying blue curve). The green box shows the common interval used for the cross-calibration (Figure 14).

### 3.4. Estimated Uncertainty

We'll use the standard deviation, *SD*, of the normalized counts by all observers for each year as a measure of the 'spread' or uncertainty. The *SD* has a clear solar cycle variation and is for most recent cycles around 8% of the count for each year. We calculate the average measured relative *SD* (shown as the purple curves at the bottom of the Figures above) for each solar cycle and derive an approximate analytical expression for the variation with time, Figure 16. The expression is only descriptive and has no other justification, but appears to be a useful estimate of the typical variation of the *SD* with time, at least capturing the observed leveling as off of the uncertainty as we move into more modern times. The standard error of the mean would be several times smaller than the *SD*, and the 95% confidence interval about the mean would be about twice that of the standard error, or about half of the *SD*. All this is under the usual over-generous assumptions about benign statistical properties of the underlying data.

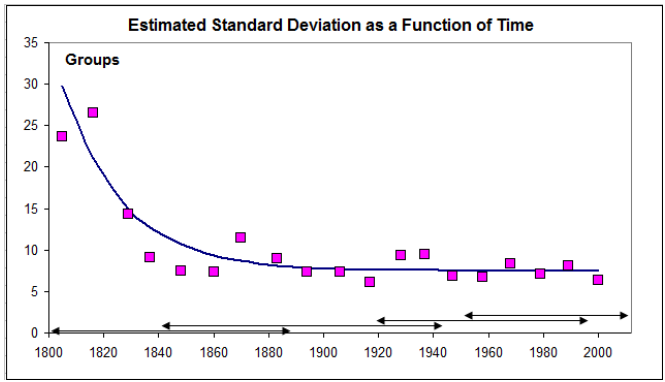

Figure 16: The average relative (in % of the group count) Standard Deviation, *SD*, computed for each solar cycle since the year 1800 and plotted at years of solar maxima. The curve is a suggested fit to the pink data points:

*SD % = 7.5+28/exp ((year-1800)/22)*

Plotting the counts by all observers for the Locarno backbone gives a visual impression of the spread of the values for a time interval where the data are supposed to be good:

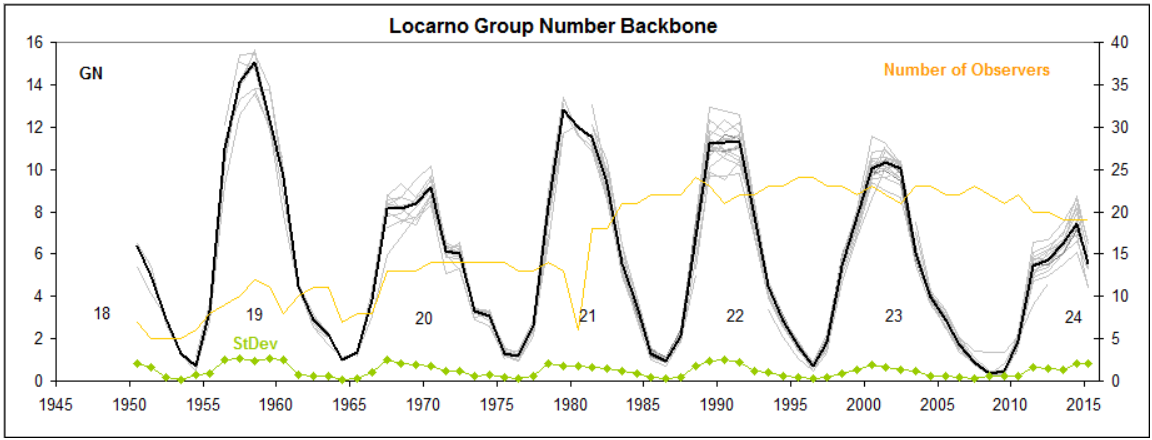

Figure 17: The Locarno backbone (heavy black line) and all the observer counts on which it is based (many light curves).

We would not expect the scatter to be *less* going back in time. HS also computed the standard deviations for their Group Sunspot Number (GSN) with the result shown in Figure 18:

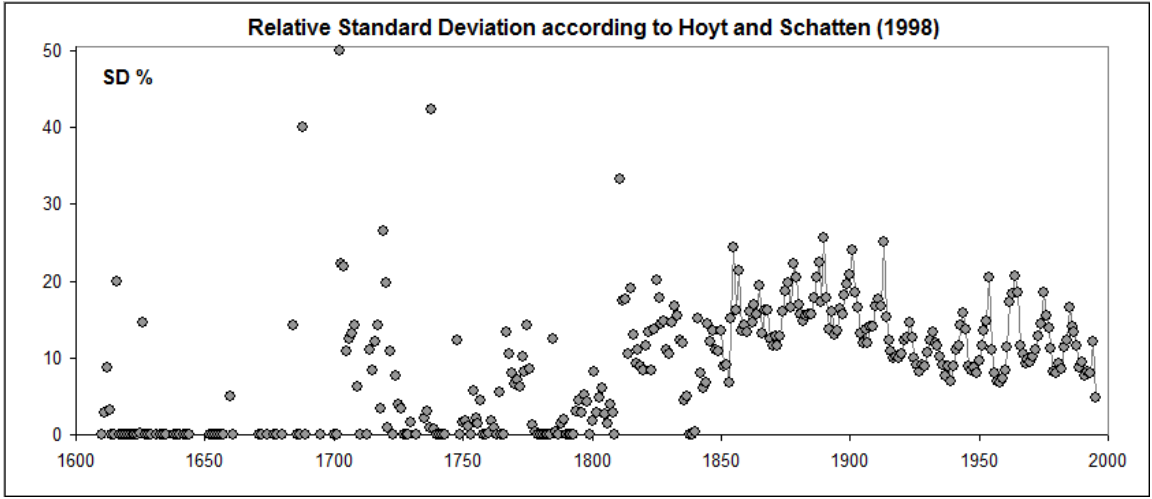

Figure 18: Relative Standard Deviations (100*SD/GSN) in percent from HS. There is a clear solar cycle variation, at least since 1850 with the relative standard deviation being twice as high during solar minima, possibly because one is dividing by a smaller number.

HS explain that they calculated "daily standard deviations of Rg for 1610 to 1995. These numbers represent the random errors in the daily means". From that, monthly and from them, the yearly means shown in Figure 18 were computed. There is possibly some confusion between the terms 'standard deviations' and 'random errors'. Elsewhere in their documentation it is stated that "the numbers provide a rough measure of the day-to-day variability of the sun each year". Whatever interpretation one prefers, it is clear that there is a dramatic change in the statistical properties of the data around 1810. We shall adopt the more realistic view that the uncertainty is higher before 1810 than later.

### 3.5. Combined Group Backbones since 1800

The scale factor between the combined Koyama-Locarno and the combined Schwabe-Wolfer backbones is 1.05±0.02. Applying this factor to the Koyama-Locarno combined backbone brings it onto the same scale as the Schwabe-Wolfer backbone that in turn is normalized to the Wolfer scale, which then becomes the 'base' scale of the whole series, Figure 19. By choosing the base scale to be in the middle of the series we minimize the accumulating effect of 'daisy-chaining' the uncertainty.

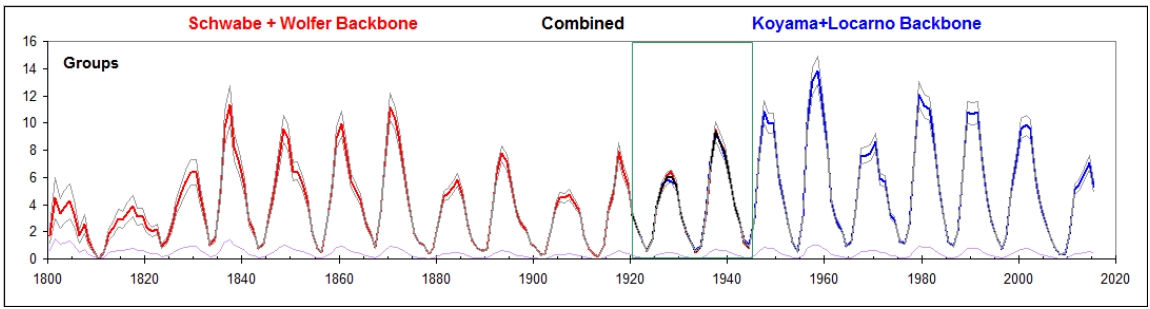

Figure 19: Group number backbone 1800-2015 derived as the mean of the composite Schwabe-Wolfer and Koyama-Locarno backbones, the latter scaled

to the former by the normalization factor 1.05±0.02, based on the common interval 1920-1944 (green box). The estimated standard deviation is shown by the purple curve at the bottom. Thin grey curves mark the ±1σ band.

### 3.6. The Staudach Backbone

Up to this point, the findings have been based on well-documented and rather plentiful data and leave very little room for valid dissent. Adding more data for the past two centuries will thus not substantially alter the conclusions one might draw from the analysis. The situation is, however, not so rosy for times before that. The primary observations during the 18[th] century must surely be the drawings of the solar disk with sunspots made by Johann Caspar Staudach (1731–1799?), digitized by Arlt (2008). As Wolf was collecting sunspot observations he became aware of a short book by Lorenz Woeckel "Die Sonne und ihre Flecken" (Woeckel, 1846) claiming that the observations by Staudach confirmed Schwabe's conclusion about a 10-year period in sunspot occurrence "as was readily seen from the (summary) table". Wolf didn't quite agree that that table with purely verbal 'evidence' was all that compelling and sought to clarify the situation by seeking access to the actual Staudach data. He was fortunate that mediation by professors Culmann and Bauernfeind resulted in obtaining in short order the original publication by the "trusting kindness" of its then owner Mr. G. Eichhorn in Nürnberg. Figure 20 shows two of the 1134 drawings (with additional verbal descriptions) made with the help of a helioscope (projection onto a sheet of paper, as in Figure 14).

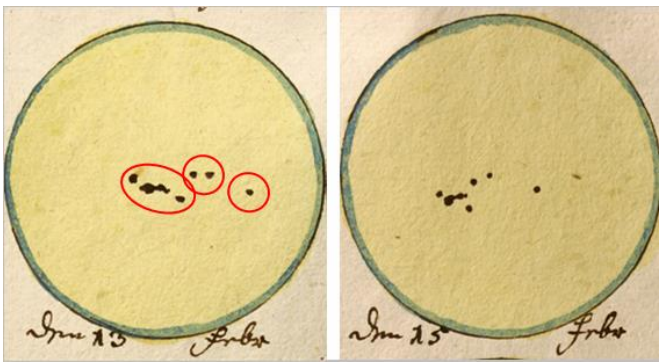

Figure 20: Staudach's drawings for the 13[th] (left) and 15[th] (right) of February 1760. On both these days Wolf (and HS) reported only a single group; a modern observer would count at least three groups on those days as indicated by the red ovals (*e.g.* S. Cortesi, Pers. Comm.). Image Courtesy R. Arlt.

Wolf determined the group and spot count for each observation using the original drawings. LS (Svalgaard 2015) has recounted the groups (and spots) for the present analysis and we'll use that recount as the base against which to normalize the counts by other observers. We found that overall, Wolf undercounted the number of groups by ~25%, compared to what we today would characterize and recognize as a 'group', while, on the other hand, we agree closely with Wolf on the spot count.

Selecting an adequate number of observers with solid overlap with Staudach is difficult as the observational material is sparse and fragmented with little assurance of homogeneity even for individual observers, and we consider the result marginal at best, Figure 21. In particular, there is very poor coverage at the end of the Staudach series, where the group counts for several consecutive years were zero and thus not amenable to determination of the scale factors needed to connect the backbone with the Schwabe backbone. Similarly, there is a dearth of observers overlapping with Staudach for the earlier years. In section 3.6.2 we shall attempt to 'bridge the gaps' in other ways.

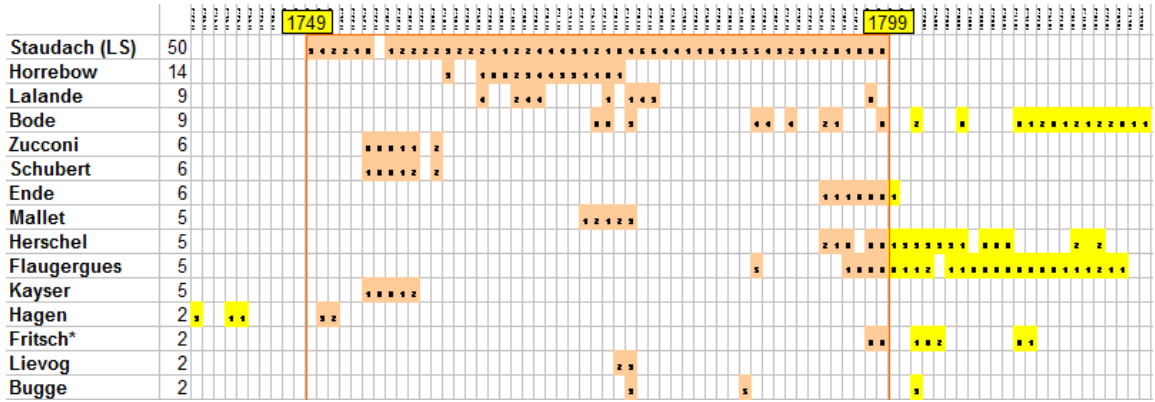

Figure 21: Coverage and observers for the Staudach backbone (1739-1822). The numbers next to the observer name signify the number of years of overlap with the primary observer, the amateur astronomer J.C. Staudach 1749-1799.

The resulting backbone, as uncertain as it is, is shown in Figure 22. The average relative standard deviation is of the order of 30%, similar to the average for the early years of the Schwabe backbone (Figure 16). We shall adopt the same 30% for the entire interval covered by the Staudach backbone.

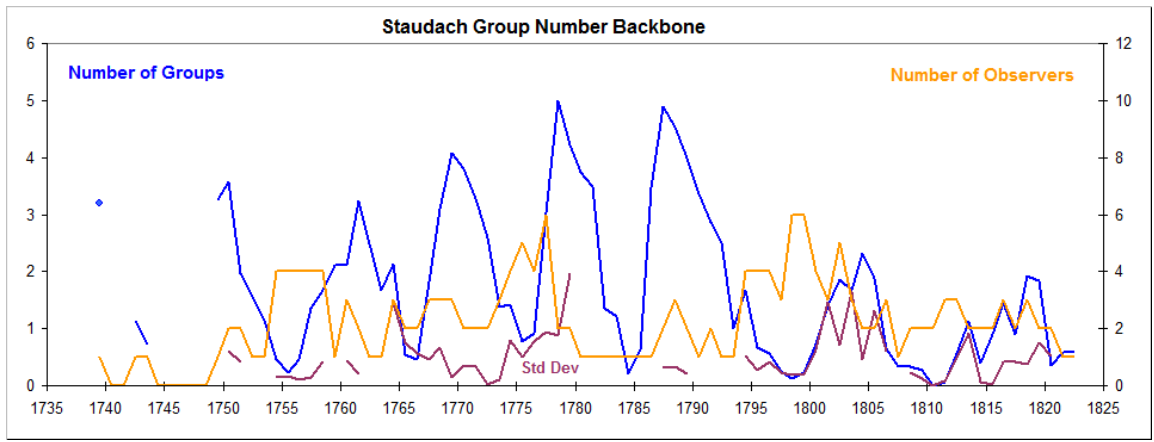

Figure 22: The Staudach backbone (blue) computed as the average of the normalized counts for the observers listed in Figure 21. The number of observers (orange) is also shown. The standard deviation of the values going into the average is shown at the bottom of the plot (purple).

### 3.6.1 Comparison with Sunspot Areas

As part of the digitization project Arlt (2008) also measured the area of the dark spots by counting pixels that were black in proportion to the total number of pixels that were part of the solar disk. A problem here was that earlier drawings showed the spots as rather bold 'blobs' in contrast to later drawings which showed more details. Figure 23 shows the variation of 183-day integration of the spot areas compared to the group counts of the backbone. The green box identifies measurements that may suffer from the over-determination (by some 30% according to Arlt) of the area possibly due to the 'boldness' of the drawings.

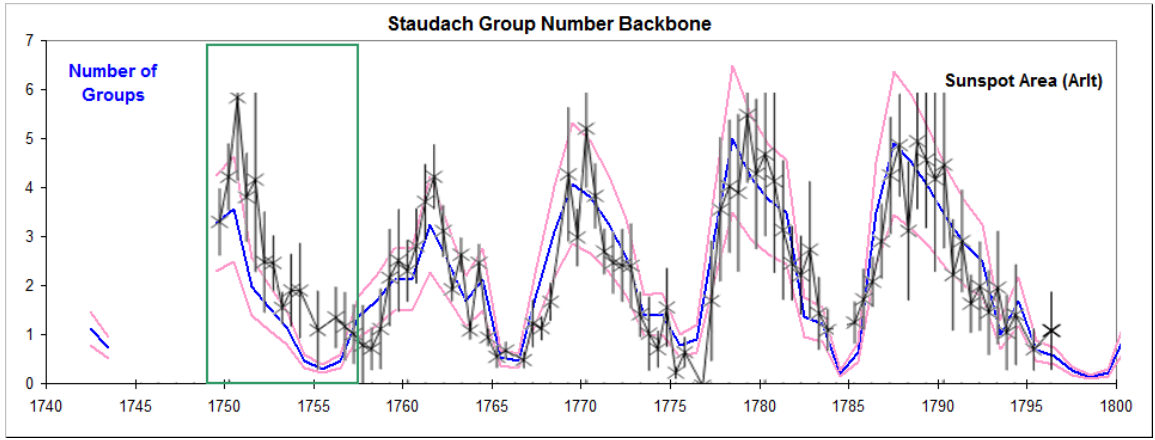

Figure 23: The Staudach backbone (blue) surrounded by the assumed ±1σ band (pink) and overlain by the sunspot areas (black) measured by Arlt (2008). The sunspot areas were integrated over ½-year intervals and are plotted with error bars computed by dividing the area by the square root of the number of observations. The area scale was set such that 1000 micro-hemispheres (μhem) correspond to 1 group, but is somewhat arbitrary, mostly determined and limited by the crudeness of the drawings.

Today, each group contributes on average 169 μhem to the total sunspot area on the disk, meaning that an area of 1000 μhem would correspond to 6 groups rather than to the 1 group on the Staudach drawings. Telescopes available to amateurs in the 18th century very likely suffered from both spherical and chromatic aberration and would not be able to show the very small groups of Waldmeier classes A and B (without penumbrae), making up about a third of all groups observed today, so that we would expect each 1000 μhem of area on Staudach's drawings to be contributing 4 = ⅔ of 6 groups to the area rather than just the single group actually observed; also the average number of spots per groups on the drawings was only 2.0, about a fifth of the typical number observed during most of modern times suggesting that the majority of the smaller spots that make up a large group were not seen by Staudach. One interpretation of this might be that the drawings are too crude (*e.g.* the spots not to scale) to allow one to establish an accurate true size of the sunspot area. At best we might hope that the *relative* sizes, e.g. of the sunspot cycle maxima, be preserved and rendered with some accuracy. Judging from Figure 23 that seems to be the case.

### 3.6.2 Bridging the Gap with Low and High Observers

Bridging the gap with poor overlap between the Schwabe and Staudach backbones calls for some innovative (and some would say perhaps slightly dubious) analysis. Examining the data for the decades surrounding the year 1800 it becomes evident that the group counts reported by the observers during that interval separate into two categories: 'low count' observers and 'high count' observers, see Table 1. It is tempting to lump together all observers in each category into two 'typical observers' for the categories. Because the categories overlap in time we can attempt to determine the scale factor needed to normalize one category to the other and in this way possibly 'bridge' the gap.

| Low Group Count Observers | | | High Group Count Observers | | |
|---|---|---|---|---|---|
| Zucconi | 1754 | 1760 | Hagen | 1739 | 1751 |
| Horrebow | 1761 | 1775 | Schubert | 1754 | 1758 |
| Mallet | 1773 | 1777 | Oriani | 1778 | 1778 |
| Lalande | 1752 | 1779 | Fritsch | 1798 | 1812 |
| Staudach | 1749 | 1799 | Lindener | 1800 | 1827 |
| Strandt | 1781 | 1799 | Herschel | 1794 | 1837 |
| Bugge | 1777 | 1810 | Prantner | 1804 | 1844 |
| Adams | 1819 | 1823 | Stark | 1813 | 1836 |
| Derfflinger | 1802 | 1824 | Tevel | 1816 | 1836 |
| Flaugergues | 1788 | 1830 | Pastorff | 1819 | 1833 |
| Arago | 1822 | 1830 | Hussey | 1826 | 1827 |
| | | | Schwarzenbrunner | 1825 | 1830 |
| | | | Schwabe | 1826 | 1857 |

Table 1: List of 'Low' observers (left) and of 'High' observers (right) with approximate range of years of observing runs used for the analysis. The two Bugge and the two Herschel observers have been placed in similar categories.

Figure 24 shows the group counts for the observers color-coded according to category:

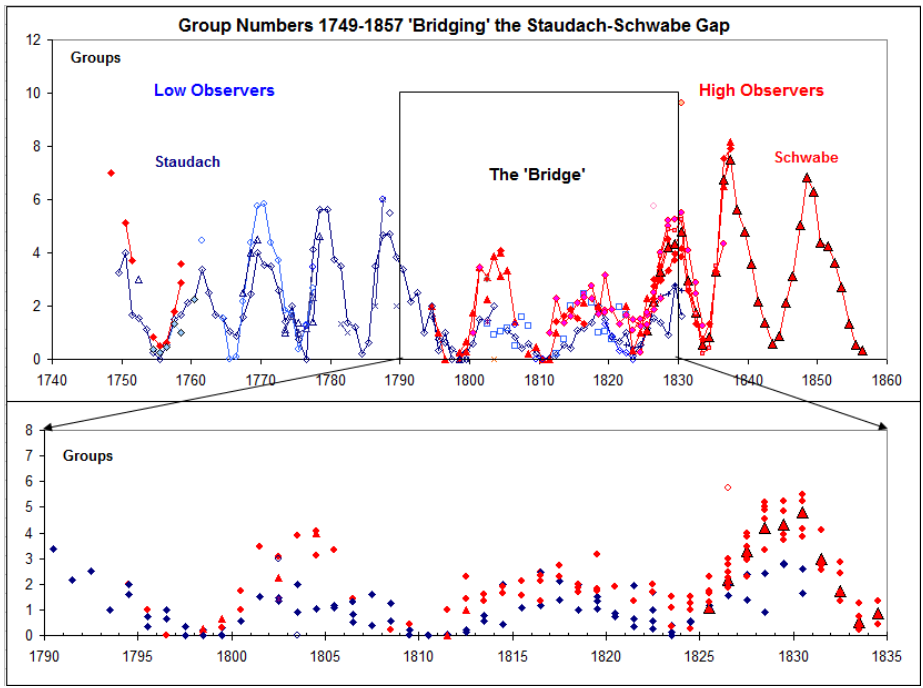

Figure 24: Un-normalized (*i.e.* raw) yearly average group counts by 'Low' observers (blue) and by 'High' observers (red). The lower panel shows an expanded view of the 'bridge' with similar color-coding.

The scale factor between the low and high categories is 1.58±0.15, so we scale the low observations to the high category by multiplying by that scale factor. Rudolf Wolf labored mightily to construct his SSN for the 18th century. It is of interest, at least for illustration, to compare his result with our 'low-high' attempt. We find, Figure 25, a satisfying agreement taking into account the difference in scales (the left-hand scale for Wolf differs from the right-hand scale for the group number by a factor of 16); apparently, we or Wolf (or both) were doing a good job.

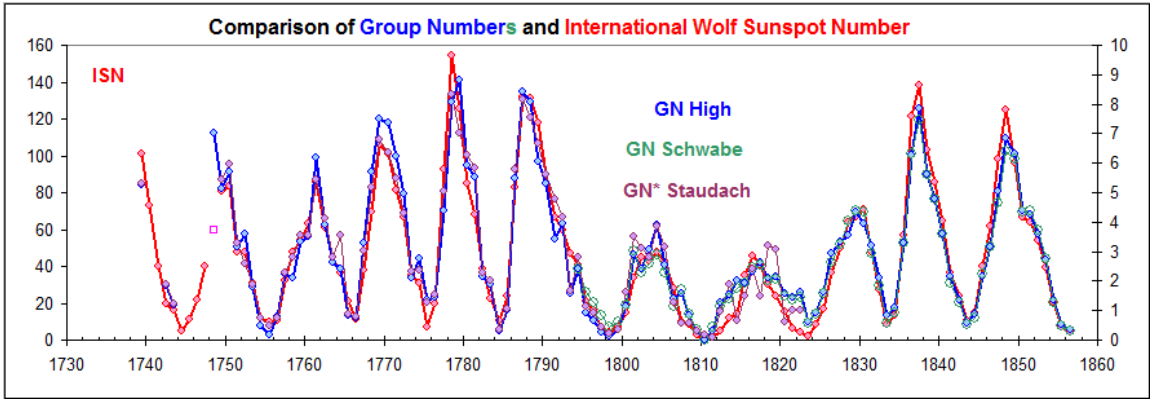

Figure 25: The International Sunspot Number constructed by Wolf and Wolfer (red curve, left-hand scale) compared to the average of the high and the scaled low Group Numbers (blue, right-hand scale). The Staudach backbone (Figure 22) scaled up by 1.68±0.05 matches the low-high Group Number scale (purple) which in turn matches the Schwabe backbone (open, green circles).

To harmonize the Staudach backbone to the base Wolfer backbone we must evidently multiply by (1.68±0.05)*(1.48±0.03) = 2.49±0.10. In Figure 26 we show the result of appending the so harmonized Staudach backbone to the composite Schwabe-Wolfer-Koyama-Locarno backbone that we derived for Figure 19:

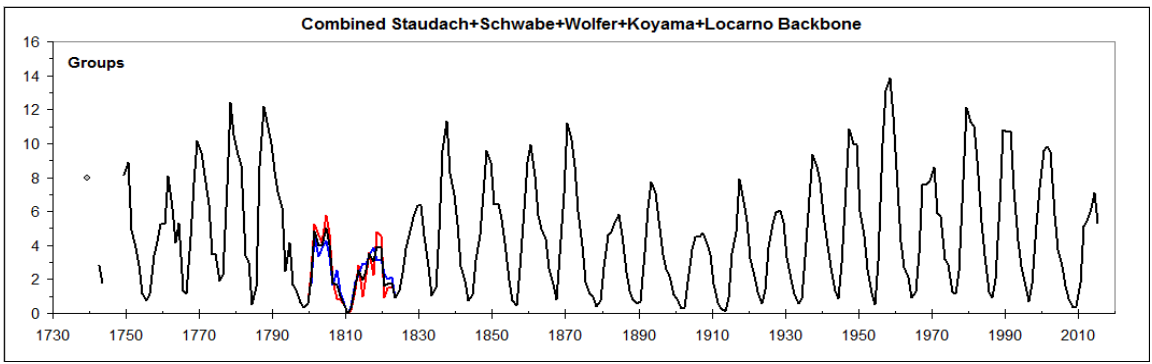

Figure 26: Combined Staudach, Schwabe, Wolfer, Koyama, and Locarno backbones, spanning from the 1730s to the present. The Staudach backbone is shown as a red curve, while the Schwabe and onwards backbone is shown as a blue curve. Where there is no overlap between the backbones only the average curve (black) is visible as the blue and red curves are covered by the average black curve.

### 3.6.3 The 'Brightest Star' Method

In Edwin Hubble's (Hubble, 1929) landmark paper showing the galaxy velocity-distance relation he used, of necessity, the brightest star in nebulae and the brightest galaxy in clusters as distance indicators, calibrated against the few nebulae whose distance could be ascertained by more reliable methods. We could apply the same procedure here and use the highest group count in each year by *any* observer as a rough indicator of solar activity (Figure 27). Hubble got his 'K-value' wrong because the first few steps of the calibration failed, teaching us the importance of those first steps and admonishing us to exercise the proverbial caution (don't we always?). An advantage of the method is that it allows us to use *all* observers without having to make a selection, although there will always be a very small number of obvious outliers to be ignored.

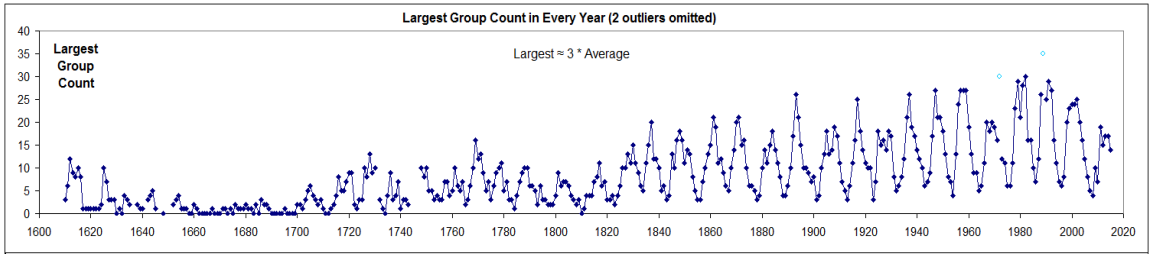

Figure 27: The largest group count by any observer during each year. That the Figure looks like Figure 1 is no accident because for this particular set of data the largest count is about 3 times the average count.

We now compare the largest group numbers with the group number from the combined backbones, seeking the scale factor for different segments of the series that give the best match between the two series, Figure 28.

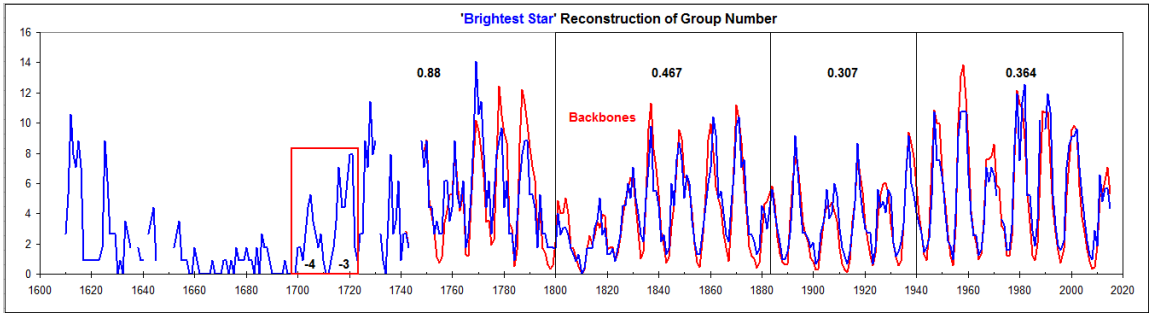

Figure 28: The largest group count by any observer during each year scaled to match the combined group number backbones. Several different scale factors are needed (given as the numbers in each of the four intervals indicated. The curve in the red box shows that solar cycles -4 and -3 around 1710 were on par with cycles around 1815 and 1910.

### 3.6.4 Sunspot Cycles -4 and -3

It is particularly noteworthy that sunspot cycles -4 and -3 (1700-1723), while small, were still in the same range as the cycles in the early part of the 20th century. Woeckel (1846) had this to say (writing before Wolf): "Um die Zeit ihrer Entdeckung waren die Flecken sehr zahlreich und man fand immer zerstreute Flecken auf der Sonne. *Scheiner* zählte ihrer oft bis 50 an einem einzigen Tage. Von 1650 bis 1670 waren sie im Gegentheil äufserst selten und nur zuweilen fand man einen oder zwei kleine Flecken. Auch von 1676 bis 1684 konnte *Flamstead* keine sehen und erst in den Jahren 1700 bis 1710 waren sie wieder sehr zahlreich." (At the time of their discovery the spots were very numerous and one could always find spots scattered over the Sun. *Scheiner* counted often up to 50 in a single day. Conversely, from 1650 to 1670 the spots were extremely rare and only occasionally could one find one or two small spots. From 1676 to 1684 *Flamsted* could also not see any spots and first in the years 1700 to 1710 were the spots again very numerous).

This is consistent with the observations of the 'red flash' (from Hα emission). It has long been known that the spicule jets move upwards along magnetic field lines rooted in the photosphere outside of sunspots, Figure 29. Thus the observation of the 'blood-red streak of light' produced by spicules requires the presence of widespread solar magnetic fields. Historical records of solar eclipse observations provide the first known report of the red flash lasting for six or seven seconds at the emersion of the sun, observed by Captain Stannyan in Bern, Switzerland, during the eclipse of 1706 (reported by Flamsted, 1708). The second observation, at the 1715 eclipse in England, was made by Edmond Halley and Jacques E.A. de Louville (Young, 1881). These observations of the red flash imply that a significant level of solar magnetism must have existed during 1706 and 1715 (Foukal and Eddy, 2007).

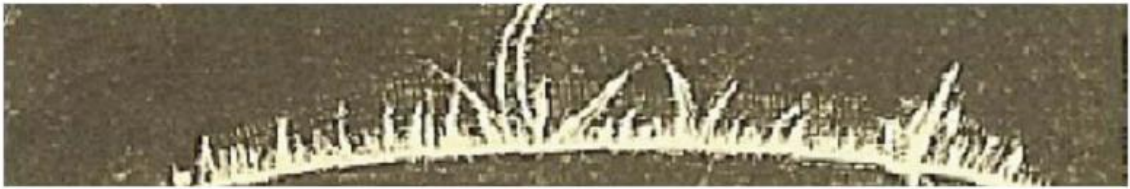

Figure 29: A drawing of the "burning prairie" appearance of the Sun's limb made by C.A. Young, on 25 July 1872. The radial structures are spicules. Drawings of these spicules by early observers led to the description of the chromosphere by the noted nineteenth-century Jesuit astronomer Angelo Secchi as a "burning prairie."

It would seem that the proper extent of the Maunder **Grand** Minimum (or at least the 'core' of the minimum, Vaquero *et al.*, 2015) should be the 40-year interval 1660-1700, rather than the commonly assumed 70-year interval 1645-1715. As already noted, all of this is contentious and currently actively researched (*e.g.* Riley *et al.*, 2015).

### 3.7. A First Estimate of the Entire Record

As all the reconstructions for times before Schwabe are highly uncertain it is not clear which one to adopt. Solar activity during those early times is an active research area and one may hope that the situation will improve. For now, we shall combine the three methods we have used, but with different (albeit somewhat arbitrary) weights: 1/1 for the Backbone, 1/2 for the Low-High method, and 1/3 for the Brightest Star method. Figure 30 shows the result:

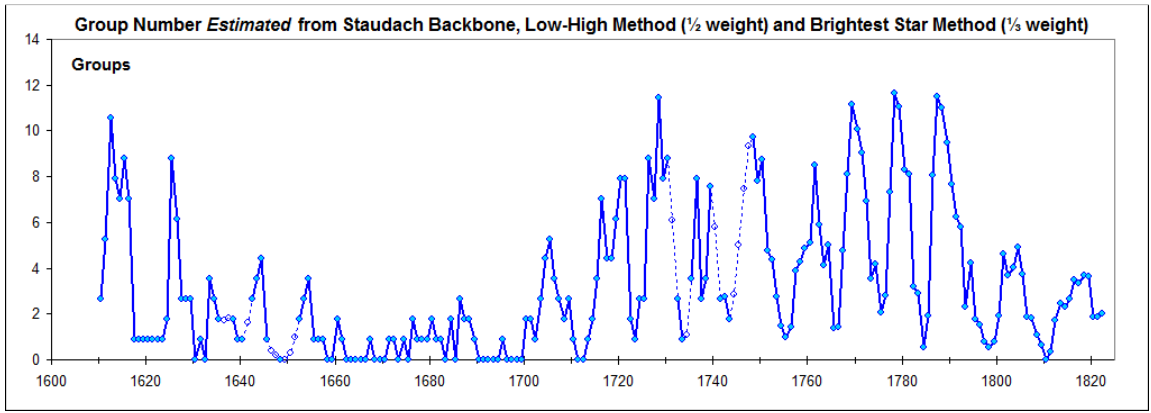

Figure 30: Yearly average Group Numbers estimated as the weighted mean of Backbones (full weight), 'Low-High' analysis (half weight), and 'Brightest Star' analysis (one-third weight). The values for years with missing data have been interpolated using third-degree polynomial fits to the surrounding years (2 years before and 2 years after the data gaps) and are plotted with open circles and dashed lines.

Combining the results of Figures 26 and 30 we get a *first* estimate of the entire 406 years of direct telescopic observations, Figure 31:

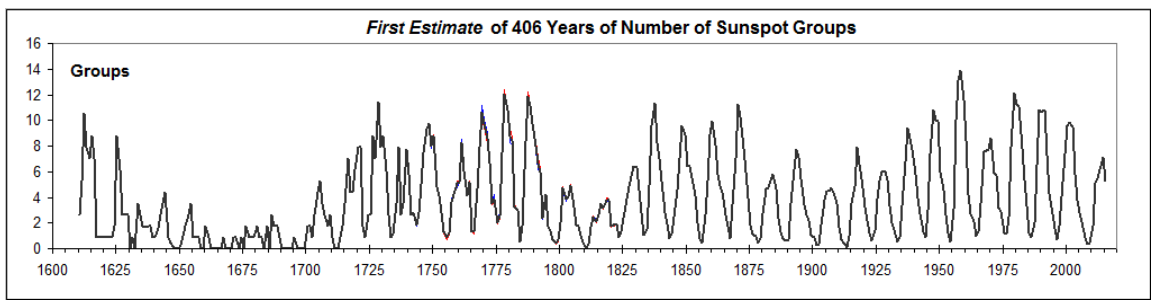

Figure 31: First estimate of the entire telescopic series. The values from Figure 26 are shown as a red curve, the values from Figure 30 as a blue curve. The black curve is the average.

## 4. What is a Sunspot Group?

Before the advent of magnetic measurements a sunspot group was defined solely on basis of its morphology and location relative to other groups. Sunspot groups were at first considered just to be spatially separate assemblies of sunspots. Schwabe's definition (1838): "Ich sehe diejenigen Fleckenhaufen als Gruppen an, die abgesondert dastehen und durch keine grösseren und kleineren Flecken und durch keinen Nebel miteinander verbunden sind." (I consider clusters of sunspots to be 'groups' if they are isolated and not connected to other clusters by larger or smaller spots or by nebulous matter). The observer Bernhard Beck (1948) describes the Zürich tradition thus "Groups are spatially isolated collections of spots. An isolated single spot also counts as a group." Friedli (2005) reminds us that after the Waldmeier Classification (1938) was introduced, the evolution of a group became a determining factor in the very definition of a group which now, in addition to be a spatially isolated collection, also must evolve as an *independent* unit, going through (at least partly) the evolutionary sequence of the Waldmeier classification, Figure 32:

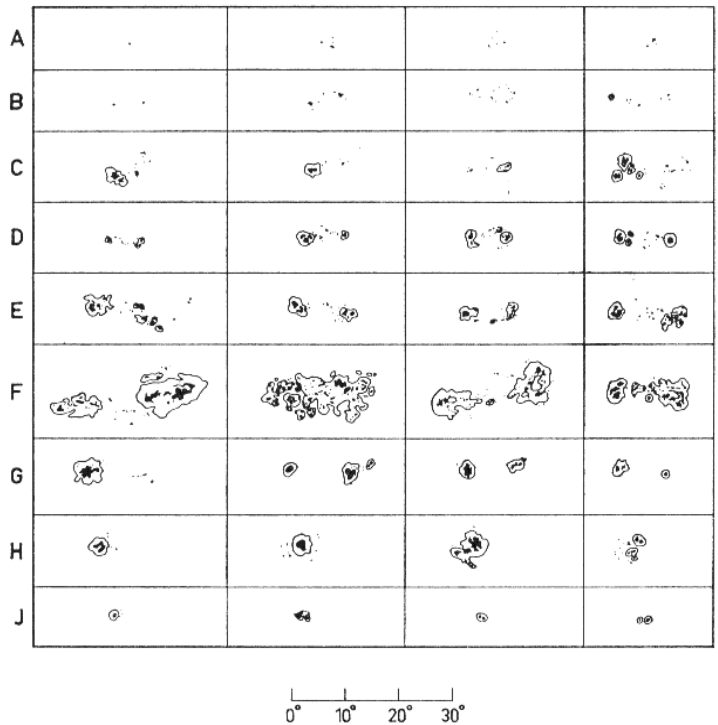

Figure 32: The Zürich or Waldmeier Classification of sunspot groups.

A: A single spot or a group of small spots without penumbra and with no bipolar structure

B: Group of spots without penumbra, but with bipolar arrangement

C: Bipolar group where one of the main spots has a penumbra

D: Bipolar group where both main spots have penumbra, but the size of the group is less that 10 °

E: Large group like D, with size at least 10 °

F: Very large and complex bipolar group with size at least 15 °

G: Bipolar group where at least one of the main spots has a penumbra, but without small spots between the main spots

H: Unipolar spot with penumbra; diameter greater than 2.5 °

J: Unipolar spot with penumbra; diameter not greater than 2.5 °

Most groups do not complete the full evolutionary sequence, and many (30-50%) do not make it past the A or B stage, but the classification is a great help in resolving and determining the group structure at times of high activity where new groups emerge next to existing groups, generally leading to an increase of the group count.

As useful as the Group Classification is, its use creates an inhomogeneity in the group count (and therefore also in the sunspot number). When an observer begins to use the evolution aspect of the Waldmeier Classification, the group count generally increases as new groups that do not fit in the evolution of existing groups are accorded an independent existence even if they morphologically would belong to an existing group, Figure 33:

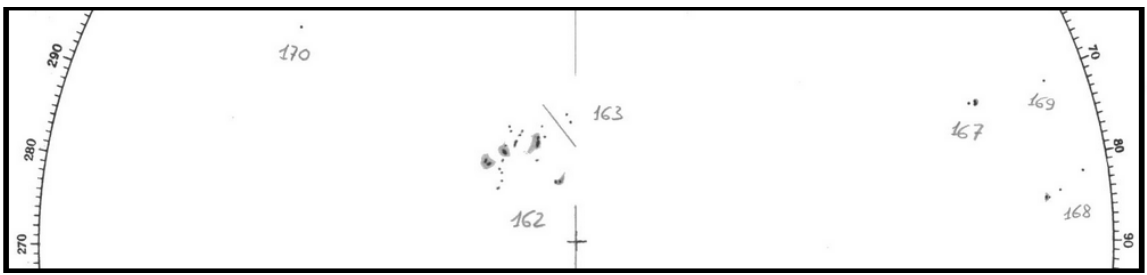

Figure 33: Typical example (from Locarno) of a group (number 163) that morphologically would be considered to be part of the larger group 162.

More examples of 'extraneous' groups (as least from a morphological viewpoint) can be seen in Figure 34. Note that the observers differ on the determination of the groups.

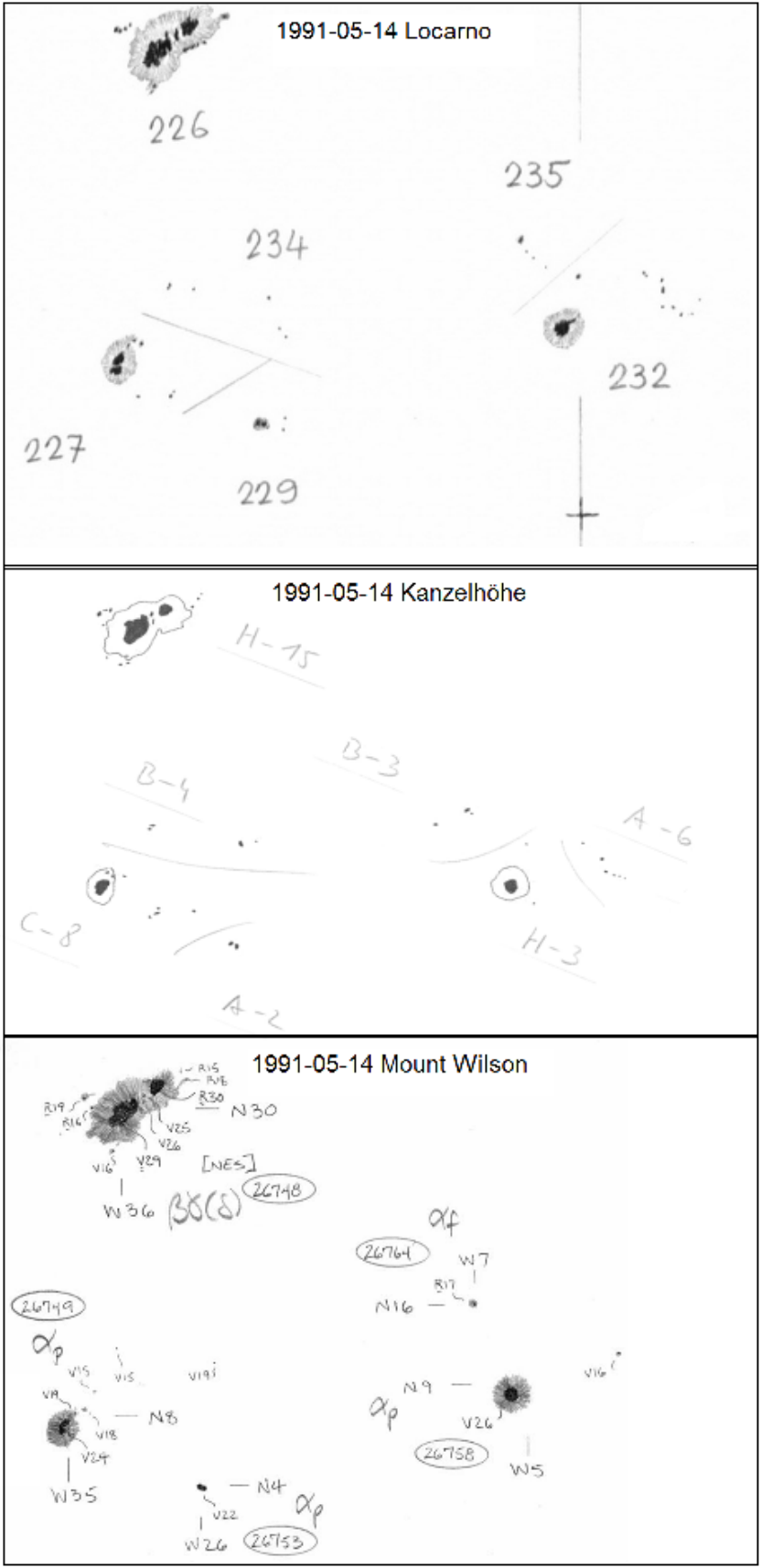

Figure 34: Drawings of the solar disk on May 14th, 1991 from three observatories (Specola Solare Ticinese, Locarno; Kanzelhöhe, Graz; and Mount Wilson, Pasadena)

The sunspot groups are as marked by the observers as well as 'demarcation' lines separating individual groups.

The middle panel illustrates the group classification (see Figure 32). The numbers next to the class letter show the number of spots in each group. The observers at Kanzelhöhe count each spot singly, regardless of size, just as Wolf and Wolfer did.

The Mount Wilson drawing includes magnetic polarities (The values V25, R19, etc. represent the spot's polarity and field strength in gauss divided by 100. The odd V, R designations come from whether the spectral line is split either to the **v**iolet end or to the **r**ed end of the spectrum. R means north polarity. V is south polarity.)

## 4.1. Effect of the Waldmeier Classification

The question now becomes: can we quantify the increase of group counts resulting from use of the evolutionary aspects of the Waldmeier Classification? One way would be to assemble the spots into groups using only the morphological rules (given in the caption to Figure 32) and not paying attention to how the groups have evolved. For the full year 2011, one of us (LS) undertook this task using the drawings made at Locarno (http://www.specola.ch/e/drawings.html) and recounted the number of groups. On the 290 drawings we find that 94 of them, or 32.4%, had at least one extra group compared to what purely morphological considerations (without evolution) would dictate. The reported average group count was 5.11, while for the groups according to the morphological rules the count was 4.77 or 7% lower. We invite the reader to repeat this exercise and get confidence in the procedure and the numbers. Taking the result that about one third of the drawings had at least one extra group (it was rare that there were more than one), we took the reported group counts on 3277 drawings for the 12-year period 2003-2014, covering a large range of solar activity, and decreased the count (if it was more than 1) by one on every third day to simulate a morphological group count. The average reported count was 3.022 versus the average 'morphological' count of 2.813, with the same 7% difference as for 2011. We therefore conclude that the effect of using the evolutionary aspects of the Waldmeier Classification is to increase the group count by 7%. We reduce the group count since 1940 by 7% to arrive at the following 'final' result, Figure 35:

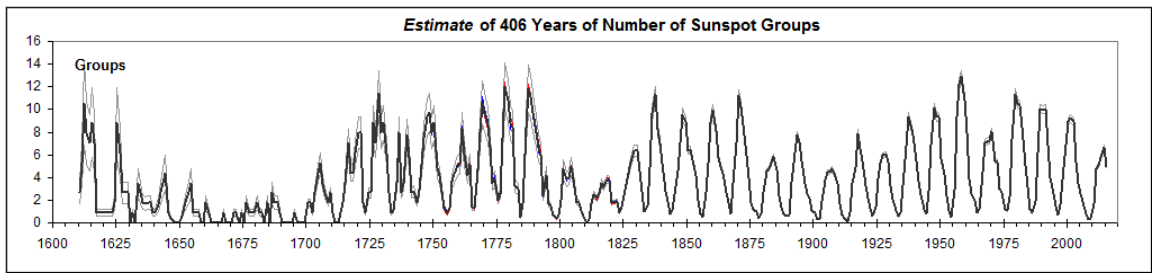

Figure 35: The estimated group count after taking into account the effect of using the evolution aspect of the Waldmeier Classification. Thin gray curves show a guesstimate of the standard error band around the mean. We used ±35% before 1700 (except for extremely low cycles where we used ±50%), half of that for 1700-1799, and half of the standard deviation suggested by Figure 16 thereafter, following the discussion in section 3.4.

For the 30 years where the group count was zero (probably because very small spots could not be observed), we have used instead a very low fiducial value of GN = 0.05 in order to avoid the problems associated with zeroes, *e.g.* in analyses involving logarithms.

Figure 36 gives an expanded view of the activity in the same format as the sunspot number graph on the SILSO website (http://www.sidc.be/images/wolfaml.png). The two graphs are similar attesting to the successful reconciliation of the group count with the 'official' International Sunspot Number. In spite of this, some notable differences are found for solar cycles -1 (~1738) and 10 (~1860), subjects for further study.

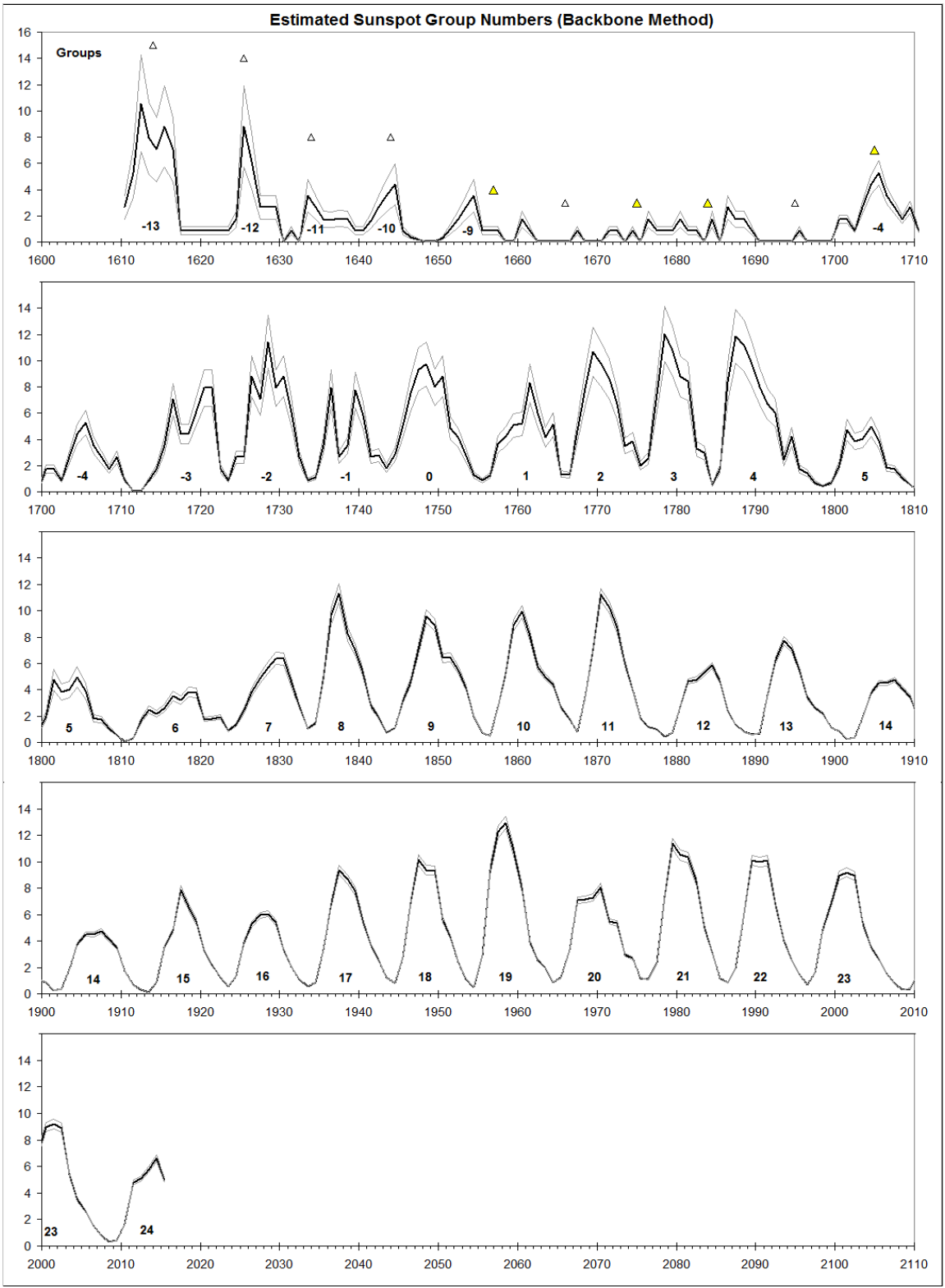

Figure 36: The estimated group count after taking into account the effect of the Waldmeier Classification. Thin gray curves outline the standard error band around the mean. Triangles show some suggested positions of solar maxima.

Of note is also that cycles -4, 5, 14, and (perhaps) 24, on average 103 years apart, are all of nearly the same amplitude; in fact, the series seems to be the 11-year Schwabe cycle amplitude modulated by a wave with period near 100 years, none of which is news. What is, perhaps, surprising is that solar activity appears to reach and sustain for extended intervals of time the same level in each of the last three centuries since 1700 and that the past several decades do not seem to have been exceptionally active, contrary to what is often claimed: "most active in 8,000 years" (*e.g.* Solanki *et al.*, 2004). That makes the core of the Maunder Minimum (1660-1700) stand out as a unique event since the telescopic observations of solar activity began.

## 5. Conclusion

We have reconstructed the sunspot group count, not by comparisons with other reconstructions and correcting those where they were deemed to be deficient, but by a re-assessment of original sources. The resulting series (a yearly summary is given in Table 2 below) is a pure solar index and does not rely on input from other proxies, *e.g.* radionuclides, auroral sightings, or geomagnetic records. Comparisons with such records are very useful and illuminating and other papers in this issue address those thoroughly. 'Backboning' the data sets, our chosen method, provides substance and rigidity by using long-time observers as a stiffness character. Our method can be easily followed and readily replicated. We have chosen to illustrate each rung on the 'sunspot scale ladder' with copious Figures as we construct the ladder, letting words accompany images, rather than the other way around. As the sunspot series (both the number of groups and the relative sunspot number) is often used by researchers in other fields for correlative studies, we have striven to make the exposition accessible to the non-specialist. It is our hope that we may succeed.

Table 2: The average Group Number (GN) for each year 1610-2015. The estimated relative ± error in % is used to indicate expected low and high limits of the Group number.

| Year | Mean GN | Low GN | High GN | Error % | Year | Mean GN | Low GN | High GN | Error % |
|---|---|---|---|---|---|---|---|---|---|
| 1610.5 | 2.64 | 1.72 | 3.56 | 35 | 1813.5 | 2.48 | 2.19 | 2.77 | 12 |
| 1611.5 | 5.28 | 3.43 | 7.13 | 35 | 1814.5 | 2.15 | 1.91 | 2.39 | 11 |
| 1612.5 | 10.56 | 6.86 | 14.26 | 35 | 1815.5 | 2.61 | 2.33 | 2.90 | 11 |
| 1613.5 | 7.92 | 5.15 | 10.69 | 35 | 1816.5 | 3.51 | 3.14 | 3.88 | 11 |
| 1614.5 | 7.04 | 4.58 | 9.50 | 35 | 1817.5 | 3.20 | 2.87 | 3.53 | 10 |
| 1615.5 | 8.80 | 5.72 | 11.88 | 35 | 1818.5 | 3.81 | 3.43 | 4.19 | 10 |
| 1616.5 | 7.04 | 4.58 | 9.50 | 35 | 1819.5 | 3.75 | 3.39 | 4.12 | 10 |
| 1617.5 | 0.88 | 0.57 | 1.19 | 35 | 1820.5 | 1.75 | 1.59 | 1.92 | 9 |
| 1618.5 | 0.88 | 0.57 | 1.19 | 35 | 1821.5 | 1.80 | 1.64 | 1.97 | 9 |
| 1619.5 | 0.88 | 0.57 | 1.19 | 35 | 1822.5 | 1.91 | 1.74 | 2.08 | 9 |
| 1620.5 | 0.88 | 0.57 | 1.19 | 35 | 1823.5 | 0.91 | 0.83 | 0.99 | 9 |
| 1621.5 | 0.88 | 0.57 | 1.19 | 35 | 1824.5 | 1.35 | 1.24 | 1.46 | 8 |
| 1622.5 | 0.88 | 0.57 | 1.19 | 35 | 1825.5 | 2.39 | 2.19 | 2.59 | 8 |
| 1623.5 | 0.88 | 0.57 | 1.19 | 35 | 1826.5 | 3.84 | 3.53 | 4.15 | 8 |
| 1624.5 | 1.76 | 1.14 | 2.38 | 35 | 1827.5 | 4.87 | 4.49 | 5.25 | 8 |
| 1625.5 | 8.80 | 5.72 | 11.88 | 35 | 1828.5 | 5.67 | 5.24 | 6.10 | 8 |
| 1626.5 | 6.16 | 4.00 | 8.32 | 35 | 1829.5 | 6.38 | 5.90 | 6.86 | 7 |
| 1627.5 | 2.64 | 1.72 | 3.56 | 35 | 1830.5 | 6.35 | 5.88 | 6.82 | 7 |
| 1628.5 | 2.64 | 1.72 | 3.56 | 35 | 1831.5 | 4.55 | 4.22 | 4.88 | 7 |
| 1629.5 | 2.64 | 1.72 | 3.56 | 35 | 1832.5 | 2.79 | 2.59 | 2.99 | 7 |
| 1630.5 | 0.05 | 0.03 | 0.08 | 50 | 1833.5 | 1.04 | 0.97 | 1.11 | 7 |
| 1631.5 | 0.88 | 0.57 | 1.19 | 35 | 1834.5 | 1.48 | 1.38 | 1.58 | 7 |
| 1632.5 | 0.05 | 0.03 | 0.08 | 50 | 1835.5 | 4.91 | 4.59 | 5.23 | 7 |
| 1633.5 | 3.52 | 2.29 | 4.75 | 35 | 1836.5 | 9.61 | 8.99 | 10.23 | 6 |
| 1634.5 | 2.64 | 1.72 | 3.56 | 35 | 1837.5 | 11.32 | 10.60 | 12.04 | 6 |
| 1635.5 | 1.76 | 1.14 | 2.38 | 35 | 1838.5 | 8.26 | 7.74 | 8.78 | 6 |
| 1636.5 | 1.70 | 1.11 | 2.30 | 35 | 1839.5 | 7.01 | 6.58 | 7.44 | 6 |
| 1637.5 | 1.80 | 1.17 | 2.43 | 35 | 1840.5 | 5.26 | 4.94 | 5.58 | 6 |
| 1638.5 | 1.76 | 1.14 | 2.38 | 35 | 1841.5 | 2.78 | 2.62 | 2.94 | 6 |
| 1639.5 | 0.88 | 0.57 | 1.19 | 35 | 1842.5 | 1.92 | 1.81 | 2.03 | 6 |
| 1640.5 | 0.88 | 0.57 | 1.19 | 35 | 1843.5 | 0.75 | 0.71 | 0.79 | 6 |
| 1641.5 | 1.60 | 1.04 | 2.16 | 35 | 1844.5 | 1.11 | 1.05 | 1.17 | 6 |
| 1642.5 | 2.64 | 1.72 | 3.56 | 35 | 1845.5 | 3.14 | 2.97 | 3.31 | 6 |
| 1643.5 | 3.52 | 2.29 | 4.75 | 35 | 1846.5 | 4.60 | 4.35 | 4.85 | 5 |
| 1644.5 | 4.40 | 2.86 | 5.94 | 35 | 1847.5 | 7.07 | 6.69 | 7.45 | 5 |
| 1645.5 | 0.88 | 0.57 | 1.19 | 35 | 1848.5 | 9.58 | 9.07 | 10.09 | 5 |
| 1646.5 | 0.40 | 0.26 | 0.54 | 35 | 1849.5 | 8.92 | 8.45 | 9.39 | 5 |
| 1647.5 | 0.20 | 0.13 | 0.27 | 35 | 1850.5 | 6.40 | 6.07 | 6.73 | 5 |
| 1648.5 | 0.05 | 0.03 | 0.08 | 50 | 1851.5 | 6.45 | 6.12 | 6.78 | 5 |
| 1649.5 | 0.05 | 0.03 | 0.08 | 50 | 1852.5 | 5.52 | 5.24 | 5.80 | 5 |
| 1650.5 | 0.30 | 0.20 | 0.41 | 35 | 1853.5 | 4.15 | 3.94 | 4.36 | 5 |
| 1651.5 | 1.00 | 0.65 | 1.35 | 35 | 1854.5 | 1.99 | 1.89 | 2.09 | 5 |
| 1652.5 | 1.76 | 1.14 | 2.38 | 35 | 1855.5 | 0.77 | 0.73 | 0.81 | 5 |
| 1653.5 | 2.64 | 1.72 | 3.56 | 35 | 1856.5 | 0.48 | 0.46 | 0.50 | 5 |
| 1654.5 | 3.52 | 2.29 | 4.75 | 35 | 1857.5 | 2.65 | 2.52 | 2.78 | 5 |
| 1655.5 | 0.88 | 0.57 | 1.19 | 35 | 1858.5 | 5.26 | 5.01 | 5.51 | 5 |
| 1656.5 | 0.88 | 0.57 | 1.19 | 35 | 1859.5 | 8.88 | 8.46 | 9.30 | 5 |
| 1657.5 | 0.88 | 0.57 | 1.19 | 35 | 1860.5 | 9.94 | 9.48 | 10.40 | 5 |
| 1658.5 | 0.05 | 0.03 | 0.08 | 50 | 1861.5 | 8.10 | 7.73 | 8.47 | 5 |
| 1659.5 | 0.05 | 0.03 | 0.08 | 50 | 1862.5 | 5.76 | 5.50 | 6.02 | 5 |
| 1660.5 | 1.76 | 1.14 | 2.38 | 35 | 1863.5 | 4.96 | 4.73 | 5.19 | 5 |
| 1661.5 | 0.88 | 0.57 | 1.19 | 35 | 1864.5 | 4.38 | 4.18 | 4.58 | 5 |
| 1662.5 | 0.05 | 0.03 | 0.08 | 50 | 1865.5 | 2.67 | 2.55 | 2.79 | 4 |
| 1663.5 | 0.05 | 0.03 | 0.08 | 50 | 1866.5 | 1.76 | 1.68 | 1.84 | 4 |
| 1664.5 | 0.05 | 0.03 | 0.08 | 50 | 1867.5 | 0.82 | 0.78 | 0.86 | 4 |
| 1665.5 | 0.05 | 0.03 | 0.08 | 50 | 1868.5 | 3.45 | 3.30 | 3.60 | 4 |
| 1666.5 | 0.05 | 0.03 | 0.08 | 50 | 1869.5 | 7.06 | 6.75 | 7.37 | 4 |
| 1667.5 | 0.88 | 0.57 | 1.19 | 35 | 1870.5 | 11.22 | 10.73 | 11.71 | 4 |

| 1668.5 | 0.05 | 0.03 | 0.08 | 50 | 1871.5 | 10.25 | 9.81 | 10.69 | 4 |
|--------|------|------|------|----|--------|-------|------|-------|---|
| 1669.5 | 0.05 | 0.03 | 0.08 | 50 | 1872.5 | 8.68 | 8.31 | 9.05 | 4 |
| 1670.5 | 0.05 | 0.03 | 0.08 | 50 | 1873.5 | 5.99 | 5.74 | 6.24 | 4 |
| 1671.5 | 0.88 | 0.57 | 1.19 | 35 | 1874.5 | 3.99 | 3.82 | 4.16 | 4 |
| 1672.5 | 0.88 | 0.57 | 1.19 | 35 | 1875.5 | 1.83 | 1.75 | 1.91 | 4 |
| 1673.5 | 0.05 | 0.03 | 0.08 | 50 | 1876.5 | 1.18 | 1.13 | 1.23 | 4 |
| 1674.5 | 0.88 | 0.57 | 1.19 | 35 | 1877.5 | 0.99 | 0.95 | 1.03 | 4 |
| 1675.5 | 0.05 | 0.03 | 0.08 | 50 | 1878.5 | 0.42 | 0.40 | 0.44 | 4 |
| 1676.5 | 1.76 | 1.14 | 2.38 | 35 | 1879.5 | 0.76 | 0.73 | 0.79 | 4 |
| 1677.5 | 0.88 | 0.57 | 1.19 | 35 | 1880.5 | 2.70 | 2.59 | 2.81 | 4 |
| 1678.5 | 0.88 | 0.57 | 1.19 | 35 | 1881.5 | 4.62 | 4.43 | 4.81 | 4 |
| 1679.5 | 0.88 | 0.57 | 1.19 | 35 | 1882.5 | 4.78 | 4.58 | 4.98 | 4 |
| 1680.5 | 1.76 | 1.14 | 2.38 | 35 | 1883.5 | 5.31 | 5.09 | 5.53 | 4 |
| 1681.5 | 0.88 | 0.57 | 1.19 | 35 | 1884.5 | 5.84 | 5.60 | 6.08 | 4 |
| 1682.5 | 0.88 | 0.57 | 1.19 | 35 | 1885.5 | 4.64 | 4.45 | 4.83 | 4 |
| 1683.5 | 0.05 | 0.03 | 0.08 | 50 | 1886.5 | 2.41 | 2.31 | 2.51 | 4 |
| 1684.5 | 1.76 | 1.14 | 2.38 | 35 | 1887.5 | 1.35 | 1.30 | 1.40 | 4 |
| 1685.5 | 0.05 | 0.03 | 0.08 | 50 | 1888.5 | 0.78 | 0.75 | 0.81 | 4 |
| 1686.5 | 2.64 | 1.72 | 3.56 | 35 | 1889.5 | 0.60 | 0.58 | 0.62 | 4 |
| 1687.5 | 1.76 | 1.14 | 2.38 | 35 | 1890.5 | 0.69 | 0.66 | 0.72 | 4 |
| 1688.5 | 1.76 | 1.14 | 2.38 | 35 | 1891.5 | 3.56 | 3.42 | 3.70 | 4 |
| 1689.5 | 0.88 | 0.57 | 1.19 | 35 | 1892.5 | 6.18 | 5.94 | 6.42 | 4 |
| 1690.5 | 0.05 | 0.03 | 0.08 | 50 | 1893.5 | 7.73 | 7.42 | 8.04 | 4 |
| 1691.5 | 0.05 | 0.03 | 0.08 | 50 | 1894.5 | 7.11 | 6.83 | 7.39 | 4 |
| 1692.5 | 0.05 | 0.03 | 0.08 | 50 | 1895.5 | 5.49 | 5.27 | 5.71 | 4 |
| 1693.5 | 0.05 | 0.03 | 0.08 | 50 | 1896.5 | 3.44 | 3.30 | 3.58 | 4 |
| 1694.5 | 0.05 | 0.03 | 0.08 | 50 | 1897.5 | 2.61 | 2.51 | 2.71 | 4 |
| 1695.5 | 0.88 | 0.57 | 1.19 | 35 | 1898.5 | 2.17 | 2.09 | 2.25 | 4 |
| 1696.5 | 0.05 | 0.03 | 0.08 | 50 | 1899.5 | 1.09 | 1.05 | 1.13 | 4 |
| 1697.5 | 0.05 | 0.03 | 0.08 | 50 | 1900.5 | 0.87 | 0.84 | 0.90 | 4 |
| 1698.5 | 0.05 | 0.03 | 0.08 | 50 | 1901.5 | 0.27 | 0.26 | 0.28 | 4 |
| 1699.5 | 0.05 | 0.03 | 0.08 | 50 | 1902.5 | 0.34 | 0.33 | 0.35 | 4 |
| 1700.5 | 1.76 | 1.45 | 2.07 | 18 | 1903.5 | 1.84 | 1.77 | 1.91 | 4 |
| 1701.5 | 1.76 | 1.45 | 2.07 | 18 | 1904.5 | 3.76 | 3.61 | 3.91 | 4 |
| 1702.5 | 0.88 | 0.73 | 1.03 | 18 | 1905.5 | 4.52 | 4.35 | 4.69 | 4 |
| 1703.5 | 2.64 | 2.18 | 3.10 | 18 | 1906.5 | 4.49 | 4.32 | 4.66 | 4 |
| 1704.5 | 4.40 | 3.63 | 5.17 | 18 | 1907.5 | 4.75 | 4.57 | 4.93 | 4 |
| 1705.5 | 5.28 | 4.36 | 6.20 | 18 | 1908.5 | 4.10 | 3.94 | 4.26 | 4 |
| 1706.5 | 3.52 | 2.90 | 4.14 | 18 | 1909.5 | 3.49 | 3.36 | 3.62 | 4 |
| 1707.5 | 2.64 | 2.18 | 3.10 | 18 | 1910.5 | 1.71 | 1.64 | 1.78 | 4 |
| 1708.5 | 1.76 | 1.45 | 2.07 | 18 | 1911.5 | 0.66 | 0.63 | 0.69 | 4 |
| 1709.5 | 2.64 | 2.18 | 3.10 | 18 | 1912.5 | 0.31 | 0.30 | 0.32 | 4 |
| 1710.5 | 0.88 | 0.73 | 1.03 | 18 | 1913.5 | 0.13 | 0.13 | 0.13 | 4 |
| 1711.5 | 0.05 | 0.04 | 0.06 | 20 | 1914.5 | 0.93 | 0.89 | 0.97 | 4 |
| 1712.5 | 0.05 | 0.04 | 0.06 | 20 | 1915.5 | 3.50 | 3.37 | 3.63 | 4 |
| 1713.5 | 0.88 | 0.73 | 1.03 | 18 | 1916.5 | 4.82 | 4.64 | 5.00 | 4 |
| 1714.5 | 1.76 | 1.45 | 2.07 | 18 | 1917.5 | 7.90 | 7.60 | 8.20 | 4 |
| 1715.5 | 3.52 | 2.90 | 4.14 | 18 | 1918.5 | 6.56 | 6.31 | 6.81 | 4 |
| 1716.5 | 7.04 | 5.81 | 8.27 | 18 | 1919.5 | 5.45 | 5.24 | 5.66 | 4 |
| 1717.5 | 4.40 | 3.63 | 5.17 | 18 | 1920.5 | 3.26 | 3.14 | 3.38 | 4 |
| 1718.5 | 4.40 | 3.63 | 5.17 | 18 | 1921.5 | 2.20 | 2.12 | 2.28 | 4 |
| 1719.5 | 6.16 | 5.08 | 7.24 | 18 | 1922.5 | 1.25 | 1.20 | 1.30 | 4 |
| 1720.5 | 7.92 | 6.53 | 9.31 | 18 | 1923.5 | 0.58 | 0.56 | 0.60 | 4 |
| 1721.5 | 7.92 | 6.53 | 9.31 | 18 | 1924.5 | 1.37 | 1.32 | 1.42 | 4 |
| 1722.5 | 1.76 | 1.45 | 2.07 | 18 | 1925.5 | 3.85 | 3.70 | 4.00 | 4 |
| 1723.5 | 0.88 | 0.73 | 1.03 | 18 | 1926.5 | 5.27 | 5.07 | 5.47 | 4 |
| 1724.5 | 2.64 | 2.18 | 3.10 | 18 | 1927.5 | 5.95 | 5.72 | 6.18 | 4 |
| 1725.5 | 2.64 | 2.18 | 3.10 | 18 | 1928.5 | 6.05 | 5.82 | 6.28 | 4 |
| 1726.5 | 8.80 | 7.26 | 10.34 | 18 | 1929.5 | 5.39 | 5.19 | 5.59 | 4 |
| 1727.5 | 7.04 | 5.81 | 8.27 | 18 | 1930.5 | 3.31 | 3.18 | 3.44 | 4 |
| 1728.5 | 11.44 | 9.44 | 13.44 | 18 | 1931.5 | 2.04 | 1.96 | 2.12 | 4 |
| 1729.5 | 7.92 | 6.53 | 9.31 | 18 | 1932.5 | 1.13 | 1.09 | 1.17 | 4 |
| 1730.5 | 8.80 | 7.26 | 10.34 | 18 | 1933.5 | 0.56 | 0.54 | 0.58 | 4 |
| 1731.5 | 6.10 | 5.03 | 7.17 | 18 | 1934.5 | 0.87 | 0.84 | 0.90 | 4 |
| 1732.5 | 2.64 | 2.18 | 3.10 | 18 | 1935.5 | 3.31 | 3.18 | 3.44 | 4 |
| 1733.5 | 0.88 | 0.73 | 1.03 | 18 | 1936.5 | 6.79 | 6.53 | 7.05 | 4 |
| 1734.5 | 1.10 | 0.91 | 1.29 | 18 | 1937.5 | 9.37 | 9.02 | 9.72 | 4 |
| 1735.5 | 3.52 | 2.90 | 4.14 | 18 | 1938.5 | 8.65 | 8.32 | 8.98 | 4 |
| 1736.5 | 7.92 | 6.53 | 9.31 | 18 | 1939.5 | 7.75 | 7.46 | 8.04 | 4 |
| 1737.5 | 2.64 | 2.18 | 3.10 | 18 | 1940.5 | 5.43 | 5.22 | 5.63 | 4 |
| 1738.5 | 3.52 | 2.90 | 4.14 | 18 | 1941.5 | 3.70 | 3.56 | 3.84 | 4 |
| 1739.5 | 7.77 | 6.41 | 9.13 | 17 | 1942.5 | 2.53 | 2.44 | 2.63 | 4 |
| 1740.5 | 5.80 | 4.79 | 6.82 | 18 | 1943.5 | 1.21 | 1.17 | 1.26 | 4 |
| 1741.5 | 2.64 | 2.18 | 3.10 | 18 | 1944.5 | 0.83 | 0.80 | 0.86 | 4 |
| 1742.5 | 2.78 | 2.29 | 3.27 | 17 | 1945.5 | 2.78 | 2.67 | 2.88 | 4 |
| 1743.5 | 1.81 | 1.49 | 2.12 | 17 | 1946.5 | 6.89 | 6.63 | 7.15 | 4 |
| 1744.5 | 2.84 | 2.34 | 3.33 | 17 | 1947.5 | 10.13 | 9.75 | 10.51 | 4 |
| 1745.5 | 5.02 | 4.14 | 5.90 | 18 | 1948.5 | 9.35 | 8.99 | 9.70 | 4 |
| 1746.5 | 7.48 | 6.17 | 8.78 | 17 | 1949.5 | 9.31 | 8.96 | 9.66 | 4 |
| 1747.5 | 9.33 | 7.70 | 10.97 | 18 | 1950.5 | 5.63 | 5.41 | 5.84 | 4 |
| 1748.5 | 9.74 | 8.03 | 11.44 | 17 | 1951.5 | 4.26 | 4.10 | 4.42 | 4 |
| 1749.5 | 7.97 | 6.58 | 9.37 | 18 | 1952.5 | 2.43 | 2.34 | 2.52 | 4 |
| 1750.5 | 8.82 | 7.27 | 10.36 | 17 | 1953.5 | 1.13 | 1.09 | 1.17 | 4 |
| 1751.5 | 4.84 | 3.99 | 5.68 | 17 | 1954.5 | 0.50 | 0.49 | 0.52 | 4 |
| 1752.5 | 4.13 | 3.40 | 4.85 | 17 | 1955.5 | 2.89 | 2.78 | 3.00 | 4 |
| 1753.5 | 2.80 | 2.31 | 3.29 | 17 | 1956.5 | 9.36 | 9.00 | 9.71 | 4 |
| 1754.5 | 1.30 | 1.07 | 1.52 | 17 | 1957.5 | 12.23 | 11.77 | 12.69 | 4 |
| 1755.5 | 0.85 | 0.70 | 1.00 | 17 | 1958.5 | 12.95 | 12.47 | 13.44 | 4 |
| 1756.5 | 1.27 | 1.05 | 1.49 | 17 | 1959.5 | 10.87 | 10.46 | 11.28 | 4 |
| 1757.5 | 3.63 | 3.00 | 4.27 | 18 | 1960.5 | 7.93 | 7.63 | 8.22 | 4 |
| 1758.5 | 4.23 | 3.49 | 4.97 | 18 | 1961.5 | 3.93 | 3.78 | 4.07 | 4 |
| 1759.5 | 5.07 | 4.19 | 5.96 | 18 | 1962.5 | 2.54 | 2.45 | 2.64 | 4 |
| 1760.5 | 5.19 | 4.28 | 6.10 | 18 | 1963.5 | 1.97 | 1.90 | 2.05 | 4 |
| 1761.5 | 8.30 | 6.85 | 9.75 | 18 | 1964.5 | 0.88 | 0.85 | 0.91 | 4 |
| 1762.5 | 6.04 | 4.98 | 7.10 | 18 | 1965.5 | 1.21 | 1.16 | 1.25 | 4 |
| 1763.5 | 4.17 | 3.44 | 4.90 | 17 | 1966.5 | 3.32 | 3.19 | 3.44 | 4 |
| 1764.5 | 5.14 | 4.24 | 6.04 | 18 | 1967.5 | 7.07 | 6.81 | 7.34 | 4 |
| 1765.5 | 1.36 | 1.12 | 1.59 | 17 | 1968.5 | 7.14 | 6.87 | 7.41 | 4 |
| 1766.5 | 1.29 | 1.06 | 1.51 | 17 | 1969.5 | 7.27 | 7.00 | 7.54 | 4 |
| 1767.5 | 4.65 | 3.83 | 5.46 | 17 | 1970.5 | 8.04 | 7.74 | 8.34 | 4 |
| 1768.5 | 7.90 | 6.52 | 9.29 | 18 | 1971.5 | 5.49 | 5.28 | 5.69 | 4 |
| 1769.5 | 10.65 | 8.79 | 12.52 | 18 | 1972.5 | 5.33 | 5.13 | 5.53 | 4 |
| 1770.5 | 9.78 | 8.07 | 11.49 | 17 | 1973.5 | 2.95 | 2.84 | 3.06 | 4 |
| 1771.5 | 8.60 | 7.10 | 10.11 | 18 | 1974.5 | 2.69 | 2.59 | 2.79 | 4 |
| 1772.5 | 6.69 | 5.52 | 7.86 | 17 | 1975.5 | 1.19 | 1.14 | 1.23 | 4 |
| 1773.5 | 3.50 | 2.89 | 4.11 | 18 | 1976.5 | 1.08 | 1.04 | 1.12 | 4 |
| 1774.5 | 3.86 | 3.18 | 4.54 | 18 | 1977.5 | 2.31 | 2.22 | 2.40 | 4 |
| 1775.5 | 2.01 | 1.66 | 2.36 | 17 | 1978.5 | 7.36 | 7.09 | 7.64 | 4 |
| 1776.5 | 2.53 | 2.09 | 2.97 | 18 | 1979.5 | 11.36 | 10.93 | 11.78 | 4 |
| 1777.5 | 7.44 | 6.14 | 8.74 | 18 | 1980.5 | 10.50 | 10.11 | 10.90 | 4 |
| 1778.5 | 12.03 | 9.92 | 14.13 | 17 | 1981.5 | 10.32 | 9.93 | 10.71 | 4 |
| 1779.5 | 10.78 | 8.90 | 12.67 | 18 | 1982.5 | 8.41 | 8.10 | 8.73 | 4 |
| 1780.5 | 8.81 | 7.27 | 10.35 | 17 | 1983.5 | 5.02 | 4.83 | 5.21 | 4 |
| 1781.5 | 8.41 | 6.94 | 9.88 | 17 | 1984.5 | 3.16 | 3.04 | 3.28 | 4 |
| 1782.5 | 3.29 | 2.71 | 3.86 | 17 | 1985.5 | 1.18 | 1.13 | 1.22 | 4 |
| 1783.5 | 2.95 | 2.44 | 3.47 | 18 | 1986.5 | 0.86 | 0.83 | 0.89 | 4 |
| 1784.5 | 0.53 | 0.44 | 0.62 | 17 | 1987.5 | 1.92 | 1.84 | 1.99 | 4 |
| 1785.5 | 1.73 | 1.43 | 2.04 | 18 | 1988.5 | 6.11 | 5.88 | 6.34 | 4 |
| 1786.5 | 8.36 | 6.90 | 9.83 | 18 | 1989.5 | 10.07 | 9.70 | 10.45 | 4 |
| 1787.5 | 11.84 | 9.77 | 13.91 | 18 | 1990.5 | 9.97 | 9.60 | 10.35 | 4 |
| 1788.5 | 11.15 | 9.20 | 13.10 | 17 | 1991.5 | 10.07 | 9.69 | 10.44 | 4 |
| 1789.5 | 9.72 | 8.02 | 11.43 | 18 | 1992.5 | 6.85 | 6.59 | 7.11 | 4 |
| 1790.5 | 8.01 | 6.61 | 9.42 | 18 | 1993.5 | 4.02 | 3.87 | 4.17 | 4 |
| 1791.5 | 6.68 | 5.51 | 7.85 | 18 | 1994.5 | 2.57 | 2.47 | 2.67 | 4 |
| 1792.5 | 6.01 | 4.95 | 7.06 | 17 | 1995.5 | 1.49 | 1.43 | 1.54 | 4 |
| 1793.5 | 2.40 | 1.98 | 2.82 | 18 | 1996.5 | 0.67 | 0.65 | 0.70 | 4 |
| 1794.5 | 4.19 | 3.46 | 4.93 | 18 | 1997.5 | 1.61 | 1.55 | 1.67 | 4 |
| 1795.5 | 1.73 | 1.42 | 2.03 | 17 | 1998.5 | 4.88 | 4.70 | 5.06 | 4 |
| 1796.5 | 1.45 | 1.19 | 1.70 | 17 | 1999.5 | 6.90 | 6.64 | 7.16 | 4 |
| 1797.5 | 0.71 | 0.59 | 0.84 | 18 | 2000.5 | 8.95 | 8.62 | 9.29 | 4 |
| 1798.5 | 0.43 | 0.36 | 0.51 | 18 | 2001.5 | 9.20 | 8.85 | 9.54 | 4 |
| 1799.5 | 0.66 | 0.54 | 0.78 | 18 | 2002.5 | 8.90 | 8.56 | 9.23 | 4 |
| 1800.5 | 2.00 | 1.65 | 2.35 | 18 | 2003.5 | 5.39 | 5.19 | 5.59 | 4 |
| 1801.5 | 4.74 | 3.92 | 5.55 | 17 | 2004.5 | 3.52 | 3.39 | 3.66 | 4 |
| 1802.5 | 3.84 | 3.20 | 4.47 | 17 | 2005.5 | 2.60 | 2.50 | 2.70 | 4 |
| 1803.5 | 4.01 | 3.37 | 4.65 | 16 | 2006.5 | 1.52 | 1.47 | 1.58 | 4 |
| 1804.5 | 4.96 | 4.20 | 5.73 | 15 | 2007.5 | 0.79 | 0.76 | 0.82 | 4 |
| 1805.5 | 3.88 | 3.30 | 4.46 | 15 | 2008.5 | 0.32 | 0.31 | 0.33 | 4 |
| 1806.5 | 1.83 | 1.57 | 2.09 | 14 | 2009.5 | 0.40 | 0.39 | 0.42 | 4 |
| 1807.5 | 1.75 | 1.51 | 2.00 | 14 | 2010.5 | 1.67 | 1.61 | 1.74 | 4 |
| 1808.5 | 1.07 | 0.93 | 1.22 | 13 | 2011.5 | 4.80 | 4.62 | 4.98 | 4 |
| 1809.5 | 0.57 | 0.49 | 0.64 | 13 | 2012.5 | 5.09 | 4.90 | 5.28 | 4 |
| 1810.5 | 0.05 | 0.04 | 0.06 | 20 | 2013.5 | 5.78 | 5.56 | 5.99 | 4 |
| 1811.5 | 0.32 | 0.28 | 0.36 | 12 | 2014.5 | 6.63 | 6.38 | 6.87 | 4 |
| 1812.5 | 1.67 | 1.47 | 1.87 | 12 | 2015.5 | 4.94 | 4.76 | 5.13 | 4 |

## Acknowledgements

We thank Ed Cliver for stimulating discussions and for his hospitality during LS's several (working) visits to NSO at Sunspot, NM. And Frederic Clette and the team at SILSO for access to their database. Rainer Arlt graciously provided access to photographs of Staudach's drawings, now held at the library of the Astrophysikalisches Institut, Potsdam. We thank the Locarno observers for their hospitality during our visits, for their effort to share their data with us, and for implementing an additional un-weighted sunspot counting procedure going forwards. This research has made use of NASA's Astrophysics Data System. LS thanks Stanford University for support.

## References

Arlt, R.: 2008, Digitization of sunspot drawings by Staudacher, in 1749-1796, *Solar Phys.* **247**, 399–410, doi:10.1007/s11207-007-9113-4

Clette, F., Svalgaard, L., Vaquero, J.M., Cliver, E.W.: 2014, Revisiting the Sunspot Number, A 400–Year Perspective on the Solar Cycle, *Space Sci. Rev.* **186**, 35–103, doi:10.1007/s11214-014-0074-2

Cliver, E.W., Clette, F., Svalgaard, L., Vaquero, J.M.: 2013, Recalibrating the Sunspot Number (SSN): The SSN Workshops, *Cent. Eur. Astrophys. Bull.* **37**(2), 401–416

Cliver, E.W., Clette F., Svalgaard L., Vaquero, J.M.: 2015, Recalibrating the Sunspot Number (SN): The 3rd and 4th SN Workshops, *Cent. Eur. Astrophys. Bull.* (submitted)

Cliver, E.W., Ling, A.: 2015, The Discontinuity in ~1885 in the Group Sunspot Number. *Solar Phys.* (this volume, submitted)

Flamsted, J.: 1708, Observations of the Solar Eclipse, May 12, 1706, *Phil. Trans. (London)* **25**, 2240

Foukal, P., Eddy, J.: 2007, Did the Sun's Prairie Ever Stop Burning? *Solar Phys.* **245**, 247–249, doi:10.1007/s11207-007-9057-8

Hoyt, D.V, Schatten, K.H.: 1998, Group sunspot numbers: a new solar activity reconstruction. *Solar. Phys.* **181**, 491–512

Hubble, E.: 1929, A relation between distance and radial velocity among extra–galactic nebulae, *PNAS* **15**(3), 168–173, doi:10.1073/pnas.15.3.168

Kopecký, M., Růžičková–Topolová, B., Kuklin, G.V.: 1980, On the relative inhomogeneity of long-term series of sunspot indices, *Bull. Astron. Inst. Czech.* **31**, 267–283

Koyama, H.: 1985, Observations of Sunspots 1947-1984, *Kawade Shobo Shinsha Publishers*, Tokyo, pp354, ISBN 4-309-25030-0

Leussu, R., Usoskin, I.G., Arlt, R., Mursula, K.: 2013, Inconsistency of the Wolf sunspot number series around 1848, *Astron. Astrophys.* **559**, A28

Riley, P., Lionello, R., Linker, J.A., Cliver, E., Balogh, A., Beer, J., Charbonneau, P., Crooker, N., deRosa, M., Lockwood, M., Owens, M., McCracken, K., Usoskin, I., Koutchmy, S.: Inferring the Structure of the Solar Corona and Inner Hemisphere During the Maunder Minimum Using Global Thermodynamic Magnetohydrodynamic Simulations, *Ap. J.* **802**, 105, doi:10.1088/0004-637X/8-2/2/105

Solanki, S.K., Usoskin, I.G., Kromer, B., Schüssler, M., Beer, J.: 2004, Unusual activity of the Sun during recent decades compared to the previous 11,000 years, Nature **431**, 1084–1087, doi:10.1038/nature02995

Svalgaard, L.: 2015, A Recount of Sunspot Groups on Historical Drawings, in preparation (this volume).

Vaquero, J.M., Trigo, R.M.: 2014, Revised Group Sunspot Number Values for 1640, 1652, and 1741, *Solar Phys.* **289**(3), 803–808, doi:10.1007/s11207-013-0360-2

Vaquero, J.M., Vazquez, M.: 2009, The Sun Recorded Through History, *Astrophys.Space Sci. Library* **361**, Springer, pp382, ISBN 978-0-387-92789-3, doi:10.1007/978-0-387-92790-9

Vaquero, J.M, Gallego, M.C., Usoskin, I.G., Kovaltsov, G.A.: 2011, Revisited sunspot data: a new scenario for the onset of the Maunder minimum. *Astrophys. J. Lett.* **731**, L24

Vaquero, J.M., Kovaltsov, G.A., Usoskin, I.G., Carrasco, V.M.S., Gallego, M.C.: 2015, Level and length of cyclic solar activity during the Maunder minimum as deduced from the active-day statistics, *Astron. Astrophys* **577**, A71

Willis, D.M., Coffey, H.E., Henwood, R., Erwin, E.H., Hoyt, D.V., Wild, M.N., Denig, W.F.: 2013, *Solar Phys.* **288**(1), 117–139. doi:10.1007/s11207-013-0311-y

Woeckel, L.: 1846, *Die Sonne und ihre Flecken*, 31pp, Verlag Friedrich Campe, Nürnberg

Young, C.A.: 1881, *The Sun*, D. Appleton, New York, 321pp, 182